\documentclass[fleqn,10pt]{wlscirep}
\usepackage[utf8]{inputenc}
\usepackage[T1]{fontenc}
\usepackage{algorithmic}
\usepackage[ruled,vlined]{algorithm2e}
\usepackage[switch]{lineno}
\usepackage{acronym}
\usepackage{amsmath}

\title{DenRAM: Neuromorphic Dendritic Architecture with RRAM for Efficient Temporal Processing with Delays}

\author[1,2$\dagger$]{Simone D'Agostino}
\author[1,2$\dagger$]{Filippo Moro}
\author[1$\dagger$]{Tristan Torchet}
\author[1]{Yi\u{g}it Demira\u{g}}
\author[2]{Laurent Grenouillet}
\author[1]{Giacomo Indiveri}
\author[2]{Elisa Vianello}
\author[1]{Melika Payvand}

\affil[1]{Institute for Neuroinformatics, University of Zurich and ETH Zurich, Zurich, Switzerland}
\affil[2]{CEA-Leti, Universit\'{e} Grenoble Alpes, Grenoble, France}

\affil[$\dagger$]{Equal contribution. Order is alphabetical.}
\graphicspath{{img/}}

\begin{abstract}
  An increasing number of neuroscience studies are highlighting the importance of spatial dendritic branching in pyramidal neurons in the brain for supporting non-linear computation through localized synaptic integration.
In particular, dendritic branches play a key role in temporal signal processing and feature detection, using coincidence detection (CD) mechanisms, made possible by the presence of synaptic delays that align temporally disparate inputs for effective integration.
Computational studies on spiking neural networks further highlight the significance of delays for CD operations, enabling spatio-temporal pattern recognition within feed-forward neural networks without the need for recurrent architectures. 
In this work, we present ``DenRAM'', the first realization of a spiking neural network with analog dendritic circuits, integrated into a 130\,nm technology node coupled with resistive memory (RRAM) technology. DenRAM's dendritic circuits use the RRAM devices to implement both delays and synaptic weights in the network. 
By configuring the RRAM devices to reproduce bio-realistic timescales, and through exploiting their heterogeneity, we experimentally demonstrate DenRAM's capability to replicate synaptic delay profiles, and efficiently implement CD for spatio-temporal pattern recognition.
To validate the architecture, we conduct comprehensive system-level simulations on two representative temporal benchmarks, highlighting DenRAM's resilience to analog hardware noise, and its superior accuracy compared to recurrent architectures with an equivalent number of parameters. 
DenRAM not only brings rich temporal processing capabilities to neuromorphic architectures, but also reduces the memory footprint of edge devices, provides high accuracy on temporal benchmarks, and represents a significant step-forward in low-power real-time signal processing technologies.
\end{abstract}
\begin{document}

\acrodef{AI}[AI]{Artificial Intelligence}
\acrodef{ADC}[ADC]{Analog to Digital Converter}
\acrodef{ADEXP}[AdExp-I\&F]{Adaptive-Exponential Integrate and Fire}
\acrodef{AER}[AER]{Address-Event Representation}
\acrodef{AEX}[AEX]{AER EXtension board}
\acrodef{AE}[AE]{Address-Event}
\acrodef{AFM}[AFM]{Atomic Force Microscope}
\acrodef{AGC}[AGC]{Automatic Gain Control}
\acrodef{AMDA}[AMDA]{AER Motherboard with D/A converters}
\acrodef{ANN}[ANN]{Artificial Neural Network}
\acrodef{API}[API]{Application Programming Interface}
\acrodef{ARM}[ARM]{Advanced RISC Machine}
\acrodef{ASIC}[ASIC]{Application Specific Integrated Circuit}
\acrodef{AdExp}[AdExp-IF]{Adaptive Exponential Integrate-and-Fire}
\acrodef{BCM}[BMC]{Bienenstock-Cooper-Munro}
\acrodef{BD}[BD]{Bundled Data}
\acrodef{BEOL}[BEOL]{Back-end of Line}
\acrodef{BG}[BG]{Bias Generator}
\acrodef{BMI}[BMI]{Brain-Machince Interface}
\acrodef{BPTT}[BPTT]{Backpropagation Through Time}
\acrodef{BTB}[BTB]{band-to-band tunnelling}
\acrodef{CAD}[CAD]{Computer Aided Design}
\acrodef{CAM}[CAM]{Content Addressable Memory}
\acrodef{CAVIAR}[CAVIAR]{Convolution AER Vision Architecture for Real-Time}
\acrodef{CA}[CA]{Cortical Automaton}
\acrodef{CCN}[CCN]{Cooperative and Competitive Network}
\acrodef{CDR}[CDR]{Clock-Data Recovery}
\acrodef{CFC}[CFC]{Current to Frequency Converter}
\acrodef{CHP}[CHP]{Communicating Hardware Processes}
\acrodef{CMIM}[CMIM]{Metal-insulator-metal Capacitor}
\acrodef{CML}[CML]{Current Mode Logic}
\acrodef{CMOL}[CMOL]{Hybrid CMOS nanoelectronic circuits}
\acrodef{CMOS}[CMOS]{Complementary Metal-Oxide-Semiconductor}
\acrodef{CNN}[CCN]{Convolutional Neural Network}
\acrodef{COTS}[COTS]{Commercial Off-The-Shelf}
\acrodef{CPG}[CPG]{Central Pattern Generator}
\acrodef{CPLD}[CPLD]{Complex Programmable Logic Device}
\acrodef{CPU}[CPU]{Central Processing Unit}
\acrodef{CSM}[CSM]{Cortical State Machine}
\acrodef{CSP}[CSP]{Constraint Satisfaction Problem}
\acrodef{CV}[CV]{Coefficient of Variation}
\acrodef{DAC}[DAC]{Digital to Analog Converter}
\acrodef{DAS}[DAS]{Dynamic Auditory Sensor}
\acrodef{DAVIS}[DAVIS]{Dynamic and Active Pixel Vision Sensor}
\acrodef{DBN}[DBN]{Deep Belief Network}
\acrodef{DFA}[DFA]{Deterministic Finite Automaton}
\acrodef{DIBL}[DIBL]{drain-induced-barrier-lowering}
\acrodef{DI}[DI]{delay insensitive}
\acrodef{DMA}[DMA]{Direct Memory Access}
\acrodef{DNF}[DNF]{Dynamic Neural Field}
\acrodef{DNN}[DNN]{Deep Neural Network}
\acrodef{DOF}[DOF]{Degrees of Freedom}
\acrodef{DPE}[DPE]{Dynamic Parameter Estimation}
\acrodef{DPI}[DPI]{Differential Pair Integrator}
\acrodef{DRRZ}[DR-RZ]{Dual-Rail Return-to-Zero}
\acrodef{DRAM}[DRAM]{Dynamic Random Access Memory}
\acrodef{DR}[DR]{Dual Rail}
\acrodef{DSP}[DSP]{Digital Signal Processor}
\acrodef{DVS}[DVS]{Dynamic Vision Sensor}
\acrodef{DYNAP}[DYNAP]{Dynamic Neuromorphic Asynchronous Processor}
\acrodef{EBL}[EBL]{Electron Beam Lithography}
\acrodef{EDVAC}[EDVAC]{Electronic Discrete Variable Automatic Computer}
\acrodef{EEG}[EEG]{Electroencephalography}
\acrodef{ECG}[ECG]{Electrocardiography}
\acrodef{EMG}[EMG]{Electromyography}
\acrodef{EIN}[EIN]{Excitatory-Inhibitory Network}
\acrodef{EM}[EM]{Expectation Maximization}
\acrodef{EPSC}[EPSC]{Excitatory Post-Synaptic Current}
\acrodef{EPSP}[EPSP]{Excitatory Post-Synaptic Potential}
\acrodef{ESN}[ESN]{Echo state Network }
\acrodef{EZ}[EZ]{Epileptogenic Zone}
\acrodef{FDSOI}[FDSOI]{Fully-Depleted Silicon on Insulator}
\acrodef{FET}[FET]{Field-Effect Transistor}
\acrodef{FFT}[FFT]{Fast Fourier Transform}
\acrodef{FI}[F-I]{Frequency-Current}
\acrodef{FPGA}[FPGA]{Field Programmable Gate Array}
\acrodef{FR}[FR]{Fast Ripple}
\acrodef{FSA}[FSA]{Finite State Automaton}
\acrodef{FSM}[FSM]{Finite State Machine}
\acrodef{GIDL}[GIDL]{gate-induced-drain-leakage}
\acrodef{GOPS}[GOPS]{Giga-Operations per Second}
\acrodef{GPU}[GPU]{Graphical Processing Unit}
\acrodef{GUI}[GUI]{Graphical User Interface}
\acrodef{HAL}[HAL]{Hardware Abstraction Layer}
\acrodef{HFO}[HFO]{High Frequency Oscillation}
\acrodef{HH}[H\&H]{Hodgkin \& Huxley}
\acrodef{HMM}[HMM]{Hidden Markov Model}
\acrodef{HCS}[HCS]{High-Conductive State}
\acrodef{HRS}[HRS]{High-Resistive State}
\acrodef{HR}[HR]{Human Readable}
\acrodef{HSE}[HSE]{Handshaking Expansion}
\acrodef{HW}[HW]{Hardware}
\acrodef{ICT}[ICT]{Information and Communication Technology}
\acrodef{IC}[IC]{Integrated Circuit}
\acrodef{IEEG}[iEEG]{intracranial electroencephalography}
\acrodef{IF2DWTA}[IF2DWTA]{Integrate \& Fire 2--Dimensional WTA}
\acrodef{IFSLWTA}[IFSLWTA]{Integrate \& Fire Stop Learning WTA}
\acrodef{IF}[I\&F]{Integrate-and-Fire}
\acrodef{IMU}[IMU]{Inertial Measurement Unit}
\acrodef{INCF}[INCF]{International Neuroinformatics Coordinating Facility}
\acrodef{INI}[INI]{Institute of Neuroinformatics}
\acrodef{IO}[I/O]{Input/Output}
\acrodef{IPSC}[IPSC]{Inhibitory Post-Synaptic Current}
\acrodef{IPSP}[IPSP]{Inhibitory Post-Synaptic Potential}
\acrodef{IP}[IP]{Intellectual Property}
\acrodef{ISI}[ISI]{Inter-Spike Interval}
\acrodef{IoT}[IoT]{Internet of Things}
\acrodef{JFLAP}[JFLAP]{Java - Formal Languages and Automata Package}
\acrodef{LEDR}[LEDR]{Level-Encoded Dual-Rail}
\acrodef{LFP}[LFP]{Local Field Potential}
\acrodef{LIF}[LIF]{Leaky Integrate and Fire}
\acrodef{LLC}[LLC]{Low Leakage Cell}
\acrodef{LNA}[LNA]{Low-Noise Amplifier}
\acrodef{LPF}[LPF]{Low Pass Filter}
\acrodef{LCS}[LCS]{Low-Conductive State}
\acrodef{LRS}[LRS]{Low-Resistive State}
\acrodef{LSM}[LSM]{Liquid State Machine}
\acrodef{LTD}[LTD]{Long Term Depression}
\acrodef{LTI}[LTI]{Linear Time-Invariant}
\acrodef{LTP}[LTP]{Long Term Potentiation}
\acrodef{LTU}[LTU]{Linear Threshold Unit}
\acrodef{LUT}[LUT]{Look-Up Table}
\acrodef{LVDS}[LVDS]{Low Voltage Differential Signaling}
\acrodef{MCMC}[MCMC]{Markov-Chain Monte Carlo}
\acrodef{MEMS}[MEMS]{Micro Electro Mechanical System}
\acrodef{MFR}[MFR]{Mean Firing Rate}
\acrodef{MIM}[MIM]{Metal Insulator Metal}
\acrodef{MLP}[MLP]{Multilayer Perceptron}
\acrodef{MOSCAP}[MOSCAP]{Metal Oxide Semiconductor Capacitor}
\acrodef{MOSFET}[MOSFET]{Metal Oxide Semiconductor Field-Effect Transistor}
\acrodef{MOS}[MOS]{Metal Oxide Semiconductor}
\acrodef{MRI}[MRI]{Magnetic Resonance Imaging}
\acrodef{NDFSM}[NDFSM]{Non-deterministic Finite State Machine} 
\acrodef{ND}[ND]{Noise-Driven}
\acrodef{NEF}[NEF]{Neural Engineering Framework}
\acrodef{NHML}[NHML]{Neuromorphic Hardware Mark-up Language}
\acrodef{NIL}[NIL]{Nano-Imprint Lithography}
\acrodef{NMDA}[NMDA]{N-Methyl-D-Aspartate}
\acrodef{NME}[NE]{Neuromorphic Engineering}
\acrodef{NN}[NN]{Neural Network}
\acrodef{NRZ}[NRZ]{Non-Return-to-Zero}
\acrodef{NSM}[NSM]{Neural State Machine}
\acrodef{OR}[OR]{Operating Room}
\acrodef{OTA}[OTA]{Operational Transconductance Amplifier}
\acrodef{PCB}[PCB]{Printed Circuit Board}
\acrodef{PCHB}[PCHB]{Pre-Charge Half-Buffer}
\acrodef{PCM}[PCM]{Phase Change Memory}
\acrodef{PE}[PE]{Processing Element}
\acrodef{PFA}[PFA]{Probabilistic Finite Automaton}
\acrodef{PFC}[PFC]{prefrontal cortex}
\acrodef{PFM}[PFM]{Pulse Frequency Modulation}
\acrodef{PR}[PR]{Production Rule}
\acrodef{PSC}[PSC]{Post-Synaptic Current}
\acrodef{PSP}[PSP]{Post-Synaptic Potential}
\acrodef{PSTH}[PSTH]{Peri-Stimulus Time Histogram}
\acrodef{QDI}[QDI]{Quasi Delay Insensitive}
\acrodef{RAM}[RAM]{Random Access Memory}
\acrodef{RDF}[RDF]{random dopant fluctuation}
\acrodef{RELU}[ReLu]{Rectified Linear Unit}
\acrodef{RLS}[RLS]{Recursive Least-Squares}
\acrodef{RMSE}[RMSE]{Root Mean Squared-Error}
\acrodef{RMS}[RMS]{Root Mean Squared}
\acrodef{RNN}[RNN]{Recurrent Neural Networks}
\acrodef{RRAM}[RRAM]{Resistive Random Access Memory}
\acrodef{RSNN}[RSNN]{Recurrent Spiking Neural Network}
\acrodef{ROLLS}[ROLLS]{Reconfigurable On-Line Learning Spiking}
\acrodef{RRAM}[RRAM]{Resistive Random Access Memory}
\acrodef{R}[R]{Ripples}
\acrodef{SAC}[SAC]{Selective Attention Chip}
\acrodef{SAT}[SAT]{Boolean Satisfiability Problem}
\acrodef{SCX}[SCX]{Silicon CorteX}
\acrodef{SD}[SD]{Signal-Driven}
\acrodef{SEM}[SEM]{Spike-based Expectation Maximization}
\acrodef{SLAM}[SLAM]{Simultaneous Localization and Mapping}
\acrodef{SNN}[SNN]{Spiking Neural Network}
\acrodef{SNR}[SNR]{Signal to Noise Ratio}
\acrodef{SOC}[SOC]{System-On-Chip}
\acrodef{SOI}[SOI]{Silicon on Insulator}
\acrodef{SP}[SP]{Separation Property}
\acrodef{SHD}[SHD]{Spiking Heidelberg Digit}
\acrodef{SRAM}[SRAM]{Static Random Access Memory}
\acrodef{SRNN}[SRNN]{Spiking Recurrent Neural Network}
\acrodef{STDP}[STDP]{Spike-Timing Dependent Plasticity}
\acrodef{STD}[STD]{Short-Term Depression}
\acrodef{STP}[STP]{Short-Term Plasticity}
\acrodef{STT-MRAM}[STT-MRAM]{Spin-Transfer Torque Magnetic Random Access Memory}
\acrodef{STT}[STT]{Spin-Transfer Torque}
\acrodef{SW}[SW]{Software}
\acrodef{TCAM}[TCAM]{Ternary Content-Addressable Memory}
\acrodef{TFT}[TFT]{Thin Film Transistor}
\acrodef{TPU}[TPU]{Tensor Processing Unit}
\acrodef{USB}[USB]{Universal Serial Bus}
\acrodef{VHDL}[VHDL]{VHSIC Hardware Description Language}
\acrodef{VLSI}[VLSI]{Very Large Scale Integration}
\acrodef{VOR}[VOR]{Vestibulo-Ocular Reflex}
\acrodef{WCST}[WCST]{Wisconsin Card Sorting Test}
\acrodef{WTA}[WTA]{Winner-Take-All}
\acrodef{XML}[XML]{eXtensible Mark-up Language}
\acrodef{CTXCTL}[CTXCTL]{CortexControl}
\acrodef{divmod3}[DIVMOD3]{divisibility of a number by three}
\acrodef{hWTA}[hWTA]{hard Winner-Take-All}
\acrodef{sWTA}[sWTA]{soft Winner-Take-All}
\acrodef{APMOM}[APMOM]{Alternate Polarity Metal On Metal}
\acrodef{SRNN}[SRNN]{Spiking Recurrent Neural Networks}
\acrodef{fMRI}[fMRI]{functional Magnetic Resonance Imaging}
\acrodef{RL}[RL]{Reinforcement Learning}
\acrodef{ES}[ES]{Evolutionary Strategies}
\acrodef{SL}[SL]{Source Line}
\acrodef{BL}[BL]{Bit Line}
\acrodef{WL}[WL]{Word Line}
\acrodef{CD}[CD]{Coincidence Detection}
\acrodef{NMDA}[NMDA]{N-methyl-D-aspartate}

\flushbottom
\maketitle
%\linenumbers
% Acronyms file

% keywords can be removed

\section{Introduction}

\begin{figure}
  \centering
  \includegraphics[width=1.0\textwidth]{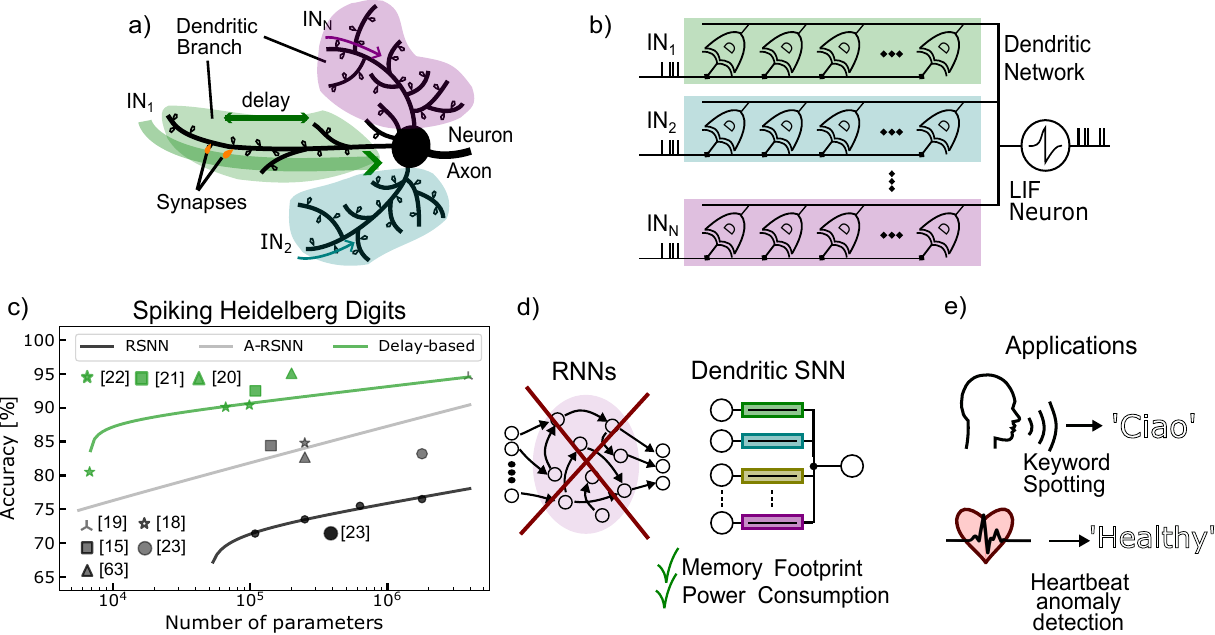}
  \caption{\textbf{Dendritic RRAM (DenRAM) concept} (a) Depiction of a biological Neuron, receiving input spikes through multiple dendritic branches. The packet of neurotransmitters travels across the dendritic branch before reaching the neuron's soma, where it is integrated. b) Scheme of the Dendritic Network, formed by several Dendritic circuits grouped into Dendritic branches macro-circuits, highlighted by different colors. The branches' outputs are integrated into a Leaky-Integrate-and-Fire Neuron. c) State-of-the-art results on the SHD dataset as a function of the number of parameters. Delay-based networks show higher accuracy and lower memory footprint compared to recurrent architectures (SRNN: recurrent spiking neural networks, A-SRNN: augmented-SRNN). d) Recurrent Neural Networks are hard to train and yield low performance. Dendritic SNNs are feed-forward models that perform better than RNNs despite reduced Memory Footprint and Power Consumption. e) Applications for the Dendritic SNN include Key-Word-Spotting and Heartbeat anomaly detection, and possibly many other sequence processing tasks.}
\label{fig:Figure1}
\end{figure}

The dendritic tree of biological neurons is an intriguing and prominent structure, with multiple branches (arbors) hosting several clusters of synapses, enabling communication and processing in complex networks. 
Communication among neurons involves the reception of action potentials (spikes) at the synapse level, generation of a post-synaptic current, with amplitude proportional to the synaptic weight, and propagation of the weighted sum of all the currents to the cell body (the soma). 
Spikes are produced in the neuron soma and transmitted through its axon to the synapses of the destination neurons (Fig.~\ref{fig:Figure1}a).
Dendritic arbors exhibit a wide range of behaviors useful for computation, such as spatio-temporal feature detection, low-pass filtering, and non-linear integration~\cite{paugam_etal_2012_SNN}. Therefore, it has been suggested that dendritic branches can be considered as semi-independent computing units in the brain\cite{poirazi2003, polsky2004,major_etal2008_NMDA,gidon_etal2020_dendritic}.
Despite these remarkable features of dendrites, most neuron models used in \acp{ANN} do not take them into account, and instead use the so-called ``point neurons'' (i.e., neurons with all synapses connected to the same node, with no spatial differentiation).
While \acp{ANN} with point-neuron models can produce remarkable results for static inputs (or for discrete sequences of static inputs), they are not ideal for processing the temporal aspects of dynamic input patterns.

A key feature of dendrites is their ability to detect local features through the spatial and temporal alignment of synapse activations within a branch, known as Coincidence Detection (CD)~\cite{antic2010, major2013}. CD can capture temporal signal features across various time scales, effectively turning dendrites into multi-time scale processing units~\cite{boahen2022dendrocentric}. The spatial arrangement of synapses on dendrites influences both local and somatic responses, with functionally related synapses forming clusters to enhance feature detection, ultimately improving computational efficiency and storage capacity~\cite{mel1993synaptic}.
The spatial arrangement of synapses on dendrites can be modeled using neuron compartments (Fig.~\ref{fig:Figure1}b), where each compartment acts as a spatio-temporal processing element.
%collectively forming a multi-time scale processor. This multi-time scale processing is achieved through synaptic delays of varying durations, and their training enables the detection of coincidences when temporal features are present. 
Spatio-temporal feature detection is found in many biological information processing circuits, such as the Barn owl's sound source localization using the Interaural Time Difference~\cite{moro2022, funabiki2011, Lazzaro_Mead89}, or the vibration source detection by the sand scorpion~\cite{brownell1979,Haessig_etal20}. These biological systems encode information in the precise timing of spike events, allowing for extremely accurate sensing at low power consumption.
Since the time of arrival of the events is at the core of this computation, temporal variables, such as delays play a significant role in computation.

Indeed, previous studies in \acp{SNN} have demonstrated that training temporal variables, such as a synapse and neuron time constant, can enhance the accuracy of network in classifying spatio-temporal patterns~\cite{yin_etal2021_SRNNAdapt, yin_etal2023_FPTTSpike, rao_etal2022_sLSTM,nowotny2022loss,bittar2022surrogate}. More recently, delays have garnered increased attention as temporal variables that enrich the computational efficiency of \acp{SNN}~\cite{hammouamri2023learning,sun2023adaptive,patino_etal2023_imec_delay}. As illustrated in Fig.~\ref{fig:Figure1}c, state-of-the-art architectures for tasks such as classifying the \ac{SHD} dataset \cite{cramer_etal2020_SHD} (a keyword spotting task) achieve the highest accuracy by incorporating delays as additional network parameters. In contrast, recurrent architectures perform poorly and require a larger number of parameters (Fig.~\ref{fig:Figure1}d). Such sensory processing tasks require delays in the time scales of 10s-100s milliseconds, and sometimes even seconds. 

To perform real-time sensory processing applications on the edge using neuromorphic \ac{SNN} hardware (Fig.~\ref{fig:Figure1}e), on-chip representation of such delay time scales is required.
Previous attempts that integrated dendritic circuits with silicon neurons in neuromorphic hardware have demonstrated their ability to detect spatio-temporal patterns in real-time and on-chip~\cite{Wang_Liu10, Huayaney_etal16, ramakrishnan_etal2013_suma, kaiser_etal2022_DenBS2, li2020power}. However, in these implementations, the computational advantages offered by the dendrites, and especially the role of delays in spatio-temporal feature detection has been limited.

Implementation of delays are costly because each synapse requires an additional memory element and a different set of network parameters. A key to this memory is its short-term dynamics, to keep the information of the incoming spike for the required amount of time.
%To implement the dendritic delays in hardware, two additional memory mechanisms are needed: a synaptic parameter that stores the duration of delay (memory complexity scales with number of synapses), and held input signals during the delay process (memory complexity scales with number of synapses and supported delay duration).
%Ideally, the goal of such an analog design is to minimize the memory footprint by implementing the delays efficiently using the analog dynamics. 
Previous implementations of on-chip delays using \ac{CMOS} technology have used digital buffers~\cite{patino_etal2023_imec_delay} or active analog circuits~\cite{sheik_etal2012_analogdelay,wang_etal2011_andredelay,Huayaney_etal16}. 
On the other hand, emerging memory technologies, e.g., \acp{RRAM}, are promising candidates for implementing these memory elements efficiently, thanks to their non-volatile,  small 3D footprint, and zero-static power properties.

While resistive memory technologies have extensively been used to implement and store the weight parameters of neural networks~\cite{Jo_etal10,Ielmini_Waser15,Serb_etal16,Li_etal18a,kingra_etal20_slim,valentian_etal2019_spirit,wan_etal2022_neurram, ambrogio_etal18equivalent, ambrogio2023, bonnet2023},  short-term dynamics~\cite{ricci2023tunable,wang2019modeling,wang_etal2017_volatileRev}, eligibility traces~\cite{demirag_etal2021_pcmtrace}, and network connectivity parameters~\cite{dalgaty_moro_demirag_etal2023_mosaic}, their use for  implementing delay elements for edge applications has not yet been fully explored. 
Recently, we have leveraged the non-volatility and controllable resistive state of \ac{RRAM} devices as a way to not only implement weights but also efficiently implement delay lines~\cite{payvand2023}. 

In this article, we present DenRAM, an RRAM-based dendritic system that has been implemented on chip. We fabricated a prototype dendritic circuit, integrating distributed weights and delay elements, utilizing a hybrid \ac{CMOS}\ac{RRAM} process. The \ac{CMOS} part of the circuit is manufactured using a low-power 130\,nm process, with the \ac{RRAM} devices fabricated on top of the \ac{CMOS} foundry layers. The RRAM technology has been specifically developed to implement both weights and delay lines in hardware. Notably, a thick 10\,nm silicon-doped Hafnium Oxide (Si:HfO) layer has been employed, resulting in a wide range of resistance values, approximately six orders of magnitude, when the device transitions from its pristine state to the low resistive state.

We experimentally demonstrate the spatio-temporal feature detection capability (i.e., CD) of DenRAM, and complement it with a novel algorithmic framework that performs the classification of sensory signals strictly using feed-forward connections. 
By explicitly representing temporal variables through the combination of RRAM-capacitor (RC) elements, DenRAM is capable of preserving temporal information without a need for recurrent connections. %\acp{RNN} keeps track of information by creating activity cycles between the neurons, which results in active power consumption on hardware. 
%The power benefits are thanks to the delays which keep the temporal information of the data in a passive fashion, without the need for active storage of data through recurrence. 
Calibrated on our experimental results, we perform hardware-aware simulations to benchmark our approach on two representative edge computing tasks, namely keyword spotting and heartbeat anomaly detection, showing how the introduction of passive delays in the network helps with reducing the memory footprint and power consumption, compared to a recurrent architecture (Fig.~\ref{fig:Figure1} d and e).

\section{Results} 
\label{sec:results}

\subsection{Hardware Measurements}

\begin{figure}[tb!]
    \centering
    \includegraphics[width=0.98\textwidth]{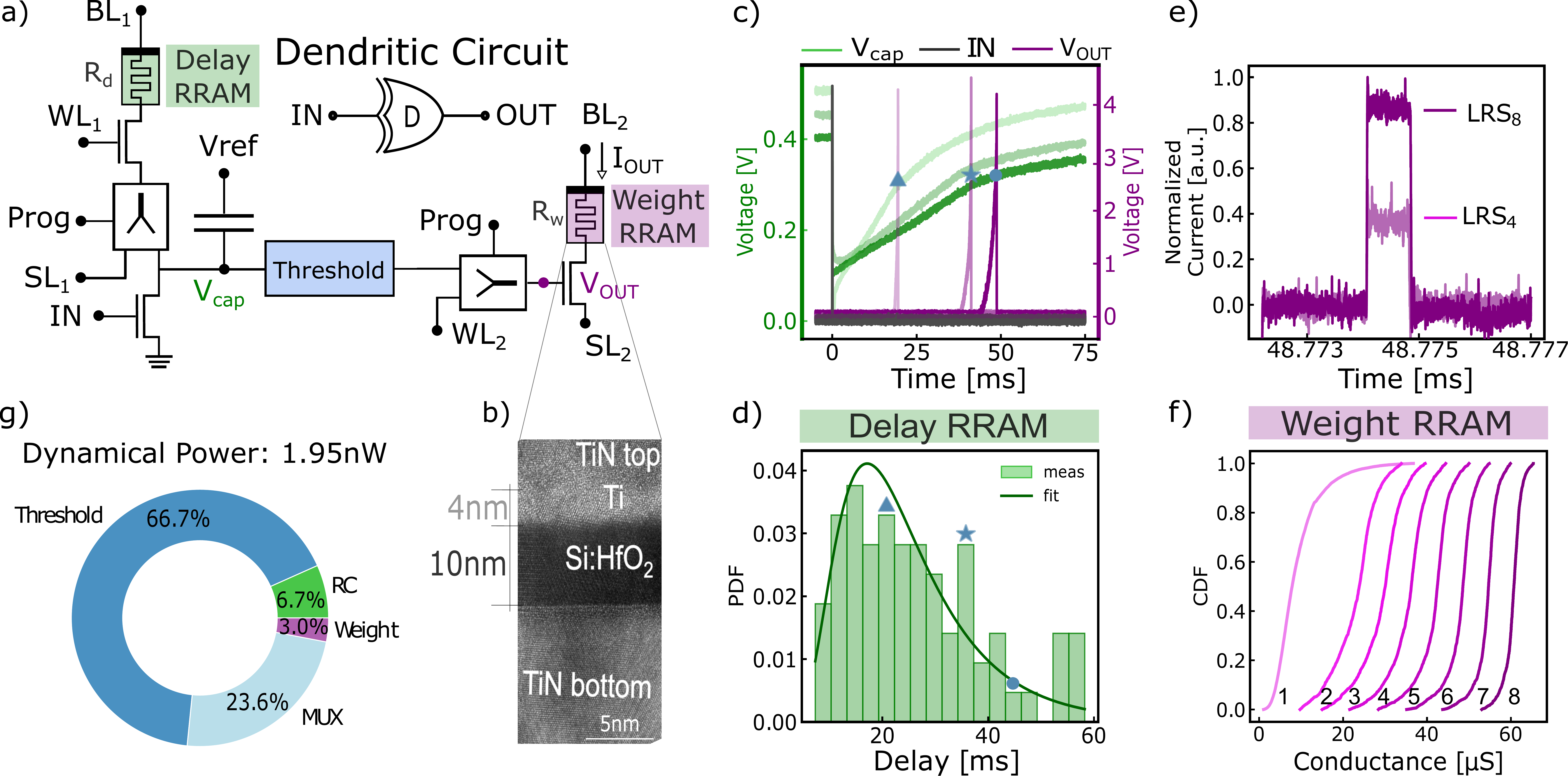}
    \caption{Dendritic circuit, the building block of the DenRAM architecture. a) Detailed schematics of the Dendritic circuit, featuring the Delay and Weight \ac{RRAM} devices, a Capacitor, dedicated multiplexers (MUX) for switching between programming and reading operations,  and a Threshold circuit. b) Scanning Electron-Microscopy image of a $Hf_{x}O$-based RRAM device used in the Dendritic circuit. c) Measurement of the Dendritic Circuit, featuring the voltage on the Input ($\mathrm{V_{IN}}$), Capacitor ($\mathrm{V_{cap}}$), and output ($\mathrm{V_{OUT}}$). d) Probability Distribution Function (PDF) of the delay measurements, with a log-normal distribution fitting curve. e) Effect of the Weight \ac{RRAM} on the output current $I_{OUT}$ measured from the Dendritic Circuit. Higher values of conductance (lower Low Resistive State) increases the output current $I_{OUT}$. f) Cumulative Distribution Function (CDF) of the Weight RRAM conductance values in their Low Resistive and High Resistive states. g) Breakdown of the dynamic power consumption of the dendritic circuit, showing the contributions from all the components in part a. The highest power is attributed to the Threshold block responsible for the 66.7\% of the total consumption. }
    \label{fig:Figure2}
\end{figure}

\paragraph*{The \ac{RRAM}-based Dendritic Circuit}
The basic building block of the DenRAM architecture is the synaptic block called ``Dendritic Circuit'', shown in Fig.~\ref{fig:Figure2}a.
It consists of a delaying unit introduced through the ($R_dC$) combination and a \emph{Threshold} unit, as well as a weighting unit through the ($R_w$) value. The delay and the weighting functionality of the synaptic circuit are both enabled by the use of two \ac{RRAM} devices per synapse. Each \ac{RRAM} is connected to its own \ac{BL}, \ac{WL}, and \ac{SL}, which are controlled to either read or program the memory device.
\ac{RRAM} devices can be either used in the reading mode, or the programming mode, through the use of the multiplexer (MUX) circuits. 
For the Delay \ac{RRAM}, the MUX \emph{Prog} selection decides whether the \ac{SL} is connected to the reading path (transistor with the gate connected to \emph{IN}), or to the programming path $SL_1$. The voltage on the $BL_1$ can be adjusted based on program or read processes. 
For the Weight \ac{RRAM}, the MUX \emph{Prog} selection decides whether the gate of the transistor is connected to the reading path from the output of the \emph{Theshold} block, or to the $WL_2$. Voltage on the $SL_2$ can be modified based on programming or reading. More details on the design of the Dendritic Circuit is provided in the Methods section.

We have built the Dendritic Circuit in a 130\,nm CMOS process, integrated with \ac{RRAM} devices in the Back End of the Line (Fig. \ref{fig:Figure2}b, more details on the Fabrication Process in the Method section). 
The fabricated devices have been thoroughly tested in their three main states (Supplementary Note 1, Fig.~S1): as fabricated, \ac{RRAM} devices are in the pristine state, where no conductive filament in oxide is present between the two electrodes, thus exhibiting the highest resistance. A one-time forming operation is carried out by applying a positive voltage across the device, causing a conductive filament to form, bringing the device to the \ac{LRS}. Then, the device becomes programmable between \ac{LRS} and a \ac{HRS}. 

%Statistical characterization of the three states of RRAM in the Dendritic Architecture is reported in Fig. \ref{fig:Figure2}e,f. 
While in the read mode, the \emph{Prog} selector is turned off (0\,V), connecting $SL_1$ to the $V_{cap}$ terminal, and the \emph{Threshold} unit output to $WL_2$. 
$BL_1$ is connected to the source voltage $V_{ref}$, and $BL_2$ is connected to the reading voltage (0.4\,V in our experiments). An input voltage pulse of 1.2\,V height, and $1\mu s$ of width is applied to the \emph{IN} terminal (Fig.~\ref{fig:Figure2}c, gray trace). This causes the \emph{IN} transistor to conduct, depolarizing the voltage on the capacitor $V_{cap}$ (Fig.~\ref{fig:Figure2}c, green trace). As soon as the \emph{IN} pulse terminates, the voltage of the capacitor recharges through the Delay \ac{RRAM}, with the time-constant $\tau = R_{d}C$, set by $R_{d}$. After some time, $V_{cap}$ crosses the threshold set by the \emph{Threshold} unit, eliciting a spike, delayed by $R_d C$ compared to $IN$, at the output of the Dendritic Circuit. 
Details on the design of the \emph{Threshold} unit is provided in Supplementary Fig.~S2a). The delayed pulse is applied to the gate of the Weight RRAM's transistor ($V_{OUT}$), allowing a current $I_{OUT}$ to flow through $R_w$.

\paragraph*{Delay characterization}
We have characterized the delay of the $R_d C$ block in 71 circuits with pristine Delay RRAMs, resulting in the distribution shown in Figure \ref{fig:Figure2}d. 
Following a log-normal distribution, the mean obtained delay is $22\,$ms, with a standard deviation of $0.5\,$ms on the underlying normal distribution.
With the conductances of our pristine devices, the minimum delay achieved is $8.08\,$ms and the maximum is $58.26\,$ms.
Such rather long delays are the result of the particularly large resistance in the Pristine state of Delay RRAMs (Figure \ref{fig:Figure2}c). Crucially, these values match the hypothesized delay produced by dendritic arbors~\cite{stuart2001, wang2000} and fall in the same order of magnitude as the temporal feature of many temporal tasks, including speech recognition ~\cite{cramer_etal2020_SHD}, biomedical signal processing~\cite{moody2001, ceolini2020hand}, and robotic control~\cite{bartolozzi2022embodied}.

\paragraph*{Weight characterization}
We provide experimental results showing the effectiveness of the Weight RRAM in modulating the output current $I_{OUT}$. As shown in Fig.~\ref{fig:Figure2}e, we measure the output current from the Dendritic Circuit in two settings.
First, we set $R_w$ to low conductance, with a weak SET operation at 1.6V, yielding a small output current. Afterward, we set the same device to higher conductance, with a strong SET pulse at 2.0V, resulting in a much larger current. As illustrated in Fig.~\ref{fig:Figure2}f, we are capable of modulating the resistance of \acp{RRAM} in a broad range of values in \ac{LRS}, enabling the modulation of the output current ($I_{OUT}$). In this case, we modulated the SET programming pulse in 8 distinct levels, obtaining 8 \ac{LRS} distributions across a 16~kbit RRAM array (refer to Supplementary Figure S1 for more details). %\YD{More quantitative characterisation of weight rram is needed e.g., more info on LRS4-8 SET pulse, we only say we program to high and low. Is Fig2f obtained with iterative programming, if so parameters etc..}

\paragraph*{Power consumption characterization}
We characterized the dynamical power consumption with thorough Spice circuit-level simulations, analyzing the contribution of the different components of the circuit (Fig.~\ref{fig:Figure2}g). In an example simulated scenario, the circuit is fed with a single input spike and produces a single output spike with 30~ms delay which is then weighted by the output RRAM device, set at $R_w=10k\Omega$. The energy produced to perform such computation is 58.5~pJ, yielding a dynamical power consumption of 1.95~nW. Of such amount, 66.7~\% is consumed in the \emph{Threshold} block, responsible for capturing the temporal evolution of the $R_d C$ circuit, and producing the delayed spike. The $R_d C$ and $R_w$ are together responsible for less than 10~\% of the power consumption. The remaining amount is attributed to the MUX selectors. A future iteration of the design will address the energy efficiency of the \emph{Threshold} block, as well as removing the necessity for the MUX selector, further improving the efficiency of the circuit.%\YD{Whats the conductance of simulated Weight RRAM? Can we also clarify what op in Threshold block consumes the most energy e.g. spike gen? I think we can also mention if there is a promsing direction to reduce power in threshold in future to be precise}

\paragraph*{The Dendritic Architecture}

\begin{figure}[t!]
    \centering
    \includegraphics[width=1\textwidth]{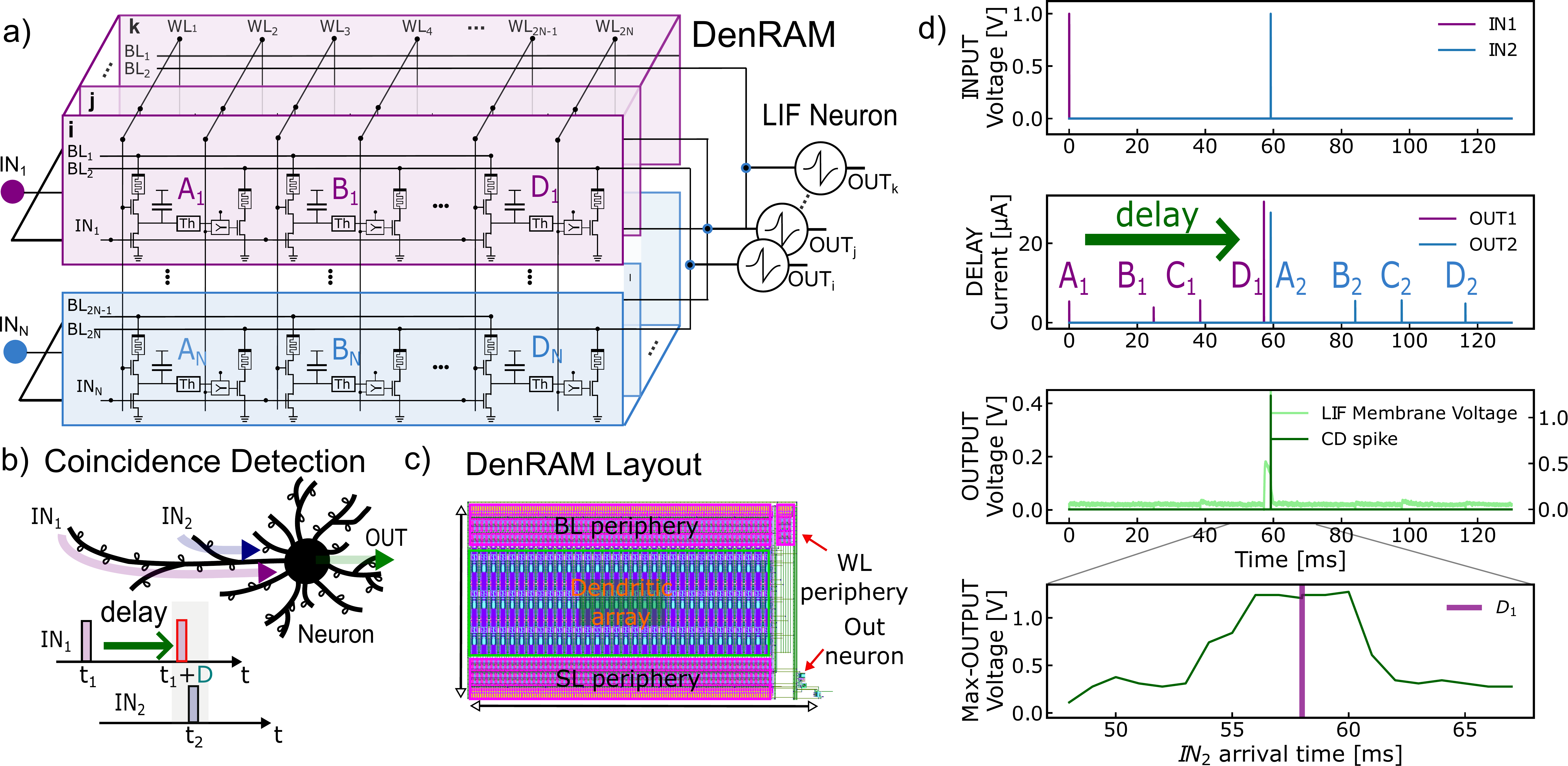}
    \caption{Coincidence Detection with Dendritic Circuits. a) Schematics of the DenRAM architecture. Two input spikes are processed by different dendritic branches, each with different set of delays. The delays that align the two input spikes receive large weights. The inputs are broadcasted to different dendritic trees leading to different output neurons. b) CD mechanism on a biological neuron, where two inputs are fed from two different synapses and reach the soma at the same time, thanks to dendritic delays. c) Physical layout of DenRAM. d) Measurement of the Dendritic Network performing coincidence detection.
    }
    \label{fig:Figure3}
\end{figure}

% neurons with different number of dendrites

The DenRAM architecture consists of an array of the Dendritic Circuits of Fig.~\ref{fig:Figure2}a, connecting to a downstream \ac{LIF} neurons, as shown in Fig.~\ref{fig:Figure3}a, featuring two of the $N$ possible dendritic branches, each having multiple synapses. Dendritic Branches can contain as many Dendritic Circuits, with the practical limit being the voltage drop as a result of the wire resistance (known as IR drop), and capacitive loading due to long metal lines (shared Source and Bit Lines). Likewise, many dendritic branches can be stacked together in parallel receiving many input channels/signals. This forms the dendritic tree of a single output neuron. Multiple dendritic trees from different output neurons can be grouped forming a layer of DenRAM that links inputs ($IN_{1,2,..., N}$) to outputs ($OUT_{i,j,...,k}$). Word Lines from all the dendritic trees can be shared in DenRAM.
%\MP{@Filippo: we need one sentence on how this will scale to multiple neurons and how the inputs will be shared. also need to change Fig. 3a to reflect this. }

The same input spike train ($IN_{i}$) is shared across a dendritic branch, which is a collection of $N$ dendritic circuits. The \ac{SL} of Weight RRAMs is shared in a branch and, when operating, is connected to the ground. 
Within a branch, the \ac{BL} of Delay RRAMs are shared, as well as the \ac{BL} of Weight \acp{RRAM}. The current from all the weight RRAMs of one branch sum with the current of the other branches, and is fed to a \ac{LIF} neuron (see Supplementary Figure~S2b for details on the \ac{LIF} circuit). As soon as the leaky integration value passes the neuron's threshold, it generates an output spike.

Therefore, the \ac{LIF} neuron receives a pre-processed version of the inputs, with a large spatio-temporal degree of freedom to adapt to a given task. 
DenRAM identifies the temporal features of the input by detecting the coinciding spikes in one branch. With each input spike on each $IN_i$ branch passing through $N$ delays, the neuron's weight parameters can be trained to select the combination of right delays, ensuring spike coincidence (Fig.~\ref{fig:Figure3}b). 
Precisely, each dendritic branch takes a spike train $x(t)=\sum_k \delta\left(t-t_k\right)$ where $t_k$ are the times at which spike occur. 
Replicates $x(t)$ $N$ times and introduces different delays, $\Delta_i$, sampled from log-normal distribution, followed by weighing with $w_i$ to obtain $S(t)=\sum_{i=1}^N w_i \cdot \sum_k \delta\left(t-\Delta_i-t_k\right)$.
It computationally performs a temporal correlation-like operation involving counting coincidences of spikes between the original and delayed trains, captured in $S(t)$. 
This architecture extends beyond the mere processing of immediate inputs; it integrates signals over a temporal spectrum, thereby establishing a dynamic form of short memory, uplifting the neuronal responsiveness to certain temporal sequences.

In real-time sensory processing settings, the temporal features in the environment are in the order of 10s to 100s of ms, which necessitates the \ac{CD}, and consequently, the delays, to be within the same timing range. This requirement enforces the use of our RRAM devices in their \ac{HRS}. However, precise control of conductances in \ac{HRS} is typically much more difficult than in \ac{LRS}, due to its high variability (Fig.~\ref{fig:Figure2}f)~\cite{esmanhotto2020}.
DenRAM takes advantage of such heterogeneity, by providing a population of analog delay elements per input channel, thus enriching the circuit with a delay spectrum that can be tuned by the weight values.
%circuit with a number of analog delays in each branch that are available as temporal variables. %In other words, each branch provides a distribution of temporal variables, i.e., delays, which could be sampled by the weight values.  
%\YD{Formulated DenRAM and rewrote 2 paragraphs above, please check}

%\MP{@Filippo: shouldnt we say here what is the size of the dendritic network that we have implemented and fabricated? A sentence that goes: we have fabricated the DenRAM of the size X by Y, whose layout is shown in Fig. 3c. }

%\MP{@Filippo, Simone: I assume the following paragraph will be re-written based on Simone's new measurements suggested by Filippo? (Sweeping the delay?)}
We perform experiments on the fabricated DenRAM (layout view in Fig~\ref{fig:Figure3}c), to showcase its temporal feature detection functionality, through \ac{CD}. The fabricated DenRAM circuit features 3 input channels or branches, 64 dendritic circuits per branch connected to a single output neuron.
The task is to detect a temporal correlation of $T\,ms$ between the spikes of two input channels. In the DenRAM scheme, this means that the delay of ~$T\,ms$ between two input channels would make the output neuron spike, and thus perform \ac{CD}. Figure~\ref{fig:Figure3}d shows the experimental measurements. 
Two inputs are presented to two different Dendritic Branches ($IN_{1}$, purple and $IN_{2}$, blue) with a temporal difference of $\sim60\,$ms (Fig~\ref{fig:Figure3}d, upper plot). 
To perform \ac{CD}, the Dendritic Network has to delay the input spike $IN_1$ by a value close to 60~ms using $R_d$, and assign a high weight to its $R_W$, so that the output \ac{LIF} neuron responds with an output spike due to the coincidence of the $IN_{2}$ with the delayed version of $IN_1$. We selected four Dendritic Circuits in the first branch ($A$ to $D$), where each of them delays the $IN_1$ spike by a given amount. Dendritic Circuit $D$ produces a delay of 58\,ms, making the delay $IN_1$ coincident with $IN_2$ from Dendritic Circuit $A$. 
Therefore, we program the RRAM Weight of the G circuit, $R_{W,G}$ to \ac{LRS} to maximize the current injected into the output neuron at the coincidence. We set all the other Dendritic Circuits' Weight RRAMs to \ac{HRS}.  
This is reflected in the measurements of the currents from the Dendritic Circuits (Fig~\ref{fig:Figure3}d, middle plot). The current from the coincident spikes of $D_1$ and $A_2$ are the highest, due to the corresponding $R_w$s at their \ac{LRS}, and the rest have lower currents.
These currents are integrated by the neuron, and its membrane voltage is measured, shown in Fig.~\ref{fig:Figure3}d, lower plot. It can be seen that the neuron responds maximally to the coincidence of spikes coming from Dendritic Circuit $D_1$ and $IN_2$, correctly performing \ac{CD}. 
To ensure the robustness of this functionality, we varied the temporal difference between the two inputs, such that the network can also separate the two coincident input spikes that are not correlated. The lower plot in Fig.~\ref{fig:Figure3}d shows the response of the output neuron when the input $IN_2$ changes its arrival time, originally at 60~ms. If $IN_2$ arrives too early or too late compared to $D_1$, the output neuron's voltage will not be fully activated, and coincidence is not detected.  An additional measurement changing the weight configuration in DenRAM is shown in Supplementary Fig. S3.

% We also tried varying the temporal difference between the two inputs verifying that coincidence was not detected in that case, ending in the demonstration of the possible feature separation of two spikes when the network evaluates them as uncorrelated (\MP{Simone:results in Supplementary Information}).

%\MP{we should remember to address the feedback of reviewers from ICONS.}

\subsection{Hardware-aware Simulations}

The biologically inspired temporal delay mechanism implemented with analog resistive memory brings superior efficiency to DenRAM.
In order to test its effectiveness on edge applications with temporally rich inputs, we benchmarked DenRAM on two sets of experiments: heartbeat anomaly detection for biomedical applications, and keyword spotting for audio processing applications.
For each benchmark, we evaluated the accuracy, memory footprint, and power consumption of DenRAM, and compared it to a conventional \ac{SRNN} with an equivalent accuracy (heartbeat task), and equivalent number of parameters (keyword spotting task). 
Both tasks require the effective learning of entangled temporal features, which is done by feed-forward synaptic weights of DenRAM that learn to utilize their corresponding synaptic delays.
%Importantly, the classification of the temporal signals are done using only feedforward connections which are enriched with delays, thanks to the DenRAM architecture. <- This is not relevant for HW.

To deliver software-comparable accuracy on the benchmarks with our analog substrate, we resorted to several hardware-aware training procedures~\cite{wan_etal2022_neurram,moro_etal2022_ISCAS} (see Methods for details).

For the weights, to emulate the limited programming resolution of RRAM devices and RRAM conductance relaxation~\cite{zhao_etal2017_filament} (which restrains the mapping of full-precision simulation weights to RRAM devices), we utilized noise-aware training~\cite{wan_etal2022_neurram} by injecting a Gaussian noise to trained network weights using the "straight-through estimator" technique.
The model of the noise is derived by conductance measurements of our programmed RRAM devices after a relaxation period (Fig.~\ref{fig:Figure2}f) and set to 10\% of the maximum conductance value in the layer~\cite{esmanhotto_etal_2022}.
This approach makes sure that the obtained simulation weights can be mappable on devices with minimal accuracy drop after realistic programming stochasticity and device relaxation effects.
% The noise represents the deviation of the conductance normalized by the maximum conductance~\cite{esmanhotto_etal_2022}.

For the delays, we used the same log-normal distribution shown in Fig.~\ref{fig:Figure2}d to sample, and then fixed the delays throughout the simulation for the heartbeat anomaly detection task. %since the achievable delays of our circuit are long enough to solve the task. 
However, our experiments with the keyword spotting task below demonstrated that speech signals require longer delays than what could be implemented with our proposed circuit (<60 ms). 
Therefore for SHD, we use a log-normal distribution with a higher mean (mean of 500\,ms for the highest accuracy) to sample and fix delays to solve the task.
This approach provides guidance for what could be achieved using higher-resistance RRAM devices in DenRAM neuromorphic architecture.

It is worth noting that in our scheme, delays are not trained and RRAM weight parameters are the only trainable parameters. 
%Rather, the mean delay is a hyperparameter that can be tuned by programming the RRAM delays, and the variability of delay RRAMs provides a rich set of delay values with that mean, from which the network can learn to select to solve the task.
Nevertheless, the network will learn to weight randomly delayed versions of the input for performing optimal \ac{CD} to detect the temporal features.  

\paragraph*{Heartbeat anomaly detection task}

\begin{figure}[tb!]
    \centering
    \includegraphics[width=1\textwidth]{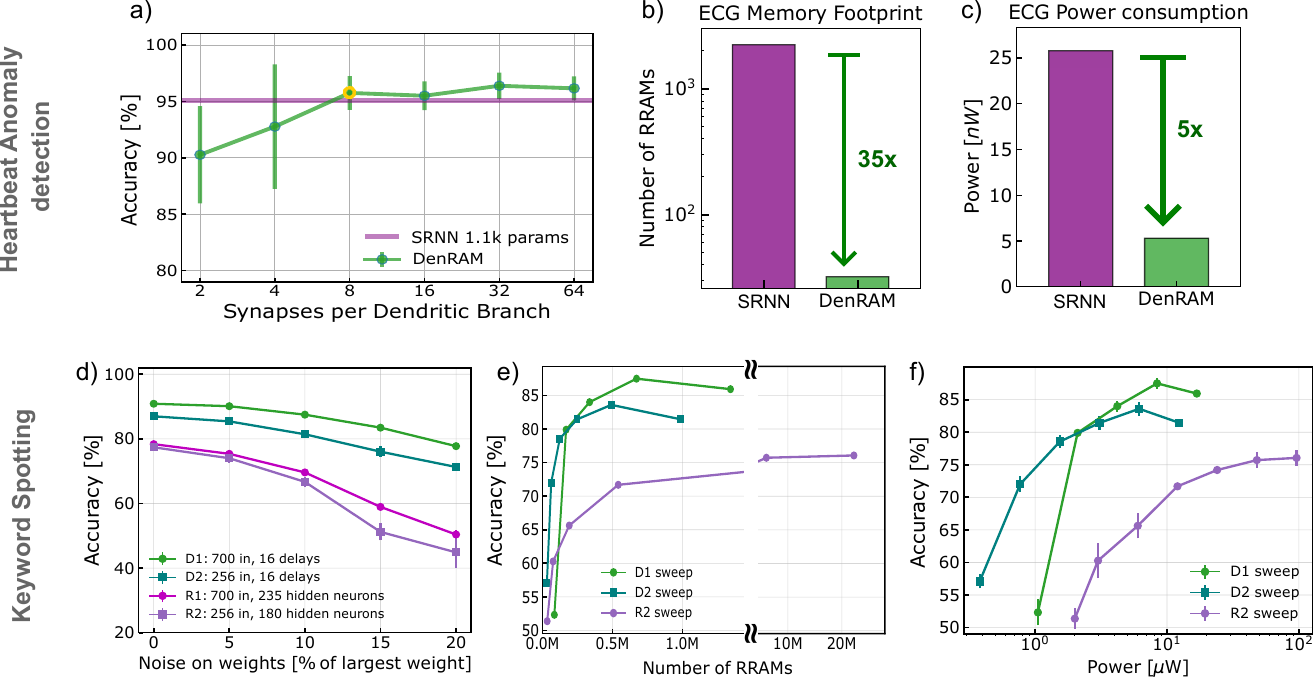}
    \caption{Performance of DenRAM on Heartbeat Anomaly Detection and Keyword Spotting. a) Classification accuracy as a function of the number of synapses per dendritic branch and comparison with a \ac{SRNN} of 32 neurons (1.1k parameters). b) Memory Footprint of the DenRAM architecture solving ECG, compared with an iso-accuracy \ac{SRNN}. c) Power consumption of DenRAM in the ECG task and comparison an iso-accuracy \ac{SRNN}. d) Classification accuracy as a function of the noise introduced on the weights for two delay architectures (D1: 700 inputs, 16 delays; D2: 256 inputs, 16 delays) and for two \ac{SRNN} architectures with one hidden layer (R1: 700 inputs, 235 hidden neurons; R2: 256 inputs, 180 hidden neurons). e) Classification accuracy (with RRAM-calibrated noise on the weights) concerning the network's number of parameters for D1, D2 and R2, sweeping the number of synapses per branch for D1 and D2, the number of hidden neurons for R2. %\TT{@Filippo: following my comment above, if we speak about the \# RRAMs the green lines should be shifted to the right to accomodate for the factor of 2. Also, "R1 sweep" should be replaced by "R2 Sweep". in the legend} 
    f) Power consumption of each network configuration (D1, D2 and R2) shown in e).} 
    \label{fig:Figure4}
\end{figure}

We first benchmark DenRAM on a heartbeat anomaly detection task using the \ac{ECG} recordings of the MIT-BIH Arrhythmia Database~\cite{moody2001}.
This is a binary classification task to distinguish between a healthy heartbeat and an arrhythmia. For the data to be compatible with DenRAM, we first encode the continuous \ac{ECG} time-series into trains of spikes using a delta-modulation technique, which describes the relative changes in signal magnitude~\cite{lee2005designing,corradi2015neuromorphic}(see Methods for details on the dataset).
This encoding scheme produces two trains of spikes for a single streaming input, one corresponding to the positive, and the other to the negative changes of the input. Therefore, the network requires only two dendritic arbors corresponding to the $IN_1$ and $IN_2$ inputs in Fig.~\ref{fig:Figure3}a.
%After the training, the inference on the test set is carried out to evaluate the performance of the architecture. 
We train the network with 10\% noise injection on the RRAM weights, sweep the number of synapses per channel, and evaluate the accuracy of the network on the test set. The results are plotted in Fig.~\ref{fig:Figure4}a. DenRAM is plotted using the green curve, which is compared against the SRNN shown in the purple curve. The error bar is variability across 5 seeds. 
%\YD{These sentences can better suit to caption of Fig4.}
DenRAM is capable of solving this temporal task with a mean accuracy of $95.30\,$\%, with 8 synapses (weights and delays) per dendritic branch, corresponding to 16 parameters and 64 devices. The network size is \textbf{35} many times smaller than the iso-accuracy SRNN made of an input layer of 2 neurons, a recurrent layer of 32 neurons, and 2 fully connected output neurons (a total of 1152 parameters, 2304 devices). It is worth noting that since the network is trained using the delay distribution of our measurements in Fig.~\ref{fig:Figure2}d, this task is solvable with a mean delay of 22\,ms. 

Avoiding the explicit recurrence in \acp{SRNN} allows DenRAM to utilize its parameters more efficiently. While the recurrent parameters in SRNNs are tuned to create complex dynamics of neuron activity, DenRAM leverages delay to explicitly process temporal features. The parameters in the recurrent layer of the SRNN scale quadratically with the number of neurons, while parameters scale linearly with input size in DenRAM. This yields a great advantage for DenRAM in the model's Memory Footprint, as highlighted in Fig.~\ref{fig:Figure4}b. 

%\MP{The folliwng paragraph is to be discussed.}
DenRAM's efficient temporal processing is also reflected in the estimation of power consumption. To estimate DenRAM's power consumption, we extend the Spice simulation presented in Fig.~\ref{fig:Figure2}g to the system level, yielding a power consumption of 5.30nW during inference. We compare DenRAM with an implementation of an iso-accuracy SRNN built with the same 130~nm RRAM-enhanced technological substrate. The iso-accuracy SRNN features 32 hidden neurons and around 1.1k parameters (see Supplementary Figure~4d). We assume efficient implementation of 
\ac{LIF}neurons (see Supplementary Figure~2b) and run Spice simulations based on this circuit. The result reveals a reduction of \textbf{5} times in power in favor of the DenRAM architecture.

\paragraph*{Keyword spotting}

We next benchmark DenRAM on a more complex task of keyword spotting using the \ac{SHD} dataset, which consists of spoken digits of 20 classes, fed through 700 Mel-spaced band-pass filters, whose output is encoded into spikes (see Methods).
To compare the generalization performance of DenRAM compared to \ac{SRNN}, we implemented four networks: D1 (DenRAM with 16 delays per channel, using 700 input channels), R1 (a \ac{SRNN} with 700 input channels and 235 hidden neurons), and their hardware-optimized counterparts, with smaller size: D2 (a network with 16 delays per channel, using 256 input channels), and R2 (a \ac{SRNN} with 256 input channels and 180 hidden neurons). For the details of the sub-sampling of D2 and R2, refer to the Methods. D1/R1 and D2/R2  have the same number of trainable parameters: 224k in the former case, and  82k in the latter case. %\TT{@Filippo: This isn't true. D1/R1 (and D2/R2) have the same number of weights (trainable parameters), but in D1 every RRAM weight is coupled with a RRAM delay so D1 has twice more RRAMs than R1. This also affects fig4e).}
Each trainable parameter translates to two RRAM devices to implement the weight (positive and negative weights) in both DenRAM and SRNN, while DenRAM also features an additional Delay RRAM per dendritic circuit, which is not a trainable parameter. 
We first compare the four networks in their resilience against the analog RRAM noise. Fig.~\ref{fig:Figure4}d shows that DenRAM architectures of D1 and D2 are consistently better performing compared to the \acp{SRNN} of R1 and R2, for any amount of noise injected into the network. D1 is the best-performing architecture with 90.88\% without noise, and is resilient to noise, with only 3.3\% of drop in accuracy for up to 10\% of injected noise on the RRAM weights, compatible with our hardware.  This is compared to a drop of 8.75\% in accuracy, in the case of R1 for the same amount of noise (from 78.37\% in the noise-less case).
On the smaller networks with reduced memory footprint, D2 and R2 accuracies achieve 81.43\% and 66.7\%, confirming that the delay-based architectures provide more expressive representations even with less number of parameters.

Next, we evaluate the accuracy of our networks as a function of the number of parameters in the network by sweeping the number of parameters of the D1, D2 and R2 networks, with 10\% of noise on the RRAM weights. Fig.~\ref{fig:Figure4}e shows that the DenRAM network has consistently better accuracy compared to the \ac{SRNN} architecture, for the same number of parameters. D1 accuracy reaches the highest accuracy of 87.5\% using 224k parameters (700 input channels and 16 delays per channel) while injecting 10\% of RRAM noise. To the best of our knowledge, this is the highest accuracy achieved on the \ac{SHD} dataset, taking into account the variability of the analog substrate (up to 10\%). Comparatively, R1 reaches 69.62\% accuracy, with an equivalent number of parameters and noise conditions.
We suspect this big performance drop in \ac{SRNN} might be due to the higher non-linear effects of the noise in the \ac{SRNN} as compared to the DenRAM, due to the recurrent connections.

%As shown in Fig.~\ref{fig:Figure4}e, a network using 16 delays per channel (a total of 224k parameters), named D1, reaches a test accuracy of 89.93\% while a one-hidden layer \ac{SRNN} using the same number of parameters, refered as R1, only gets 74.51\%. To ensure the portability of the network to the chip we introduce 10\% noise on the weights (see Methods) which reduces the accuracy of D1 by 3\% and R1 by 19.4\%. Taking the memory footprint into account we sample evenly the input channels to use only 256 inputs, 16 delays for D2 and 180 hidden neurons for R2. D2 and R2 respectively reach a test accuracy of 74.4\% and 64.1\% while being exposed to 10\% noise. 

Finally, we evaluate the accuracy of D1, D2, and R2 networks by varying their maximum power consumption, as determined from the power estimations presented in Fig.\ref{fig:Figure2}g. The results are depicted in Fig.\ref{fig:Figure4}f, where it is observed that the accuracy of the R2 network plateaus at approximately 76\%, regardless of increasing power consumption. In contrast, the D2 network achieves an accuracy of 83.6\% at a power consumption of 6.15\,$\mu$W, while the D1 network attains an accuracy of 87.5\% with a power consumption of 8.41\,$\mu$W.
%\YD{This Fig2g reference is vague. Its not clear how R2 power consumption relates to it and how it is calculated. Also we need 1-2 sentence not so obvious commentary after stating the results.}

\section{Discussion}

%We have introduced DenRAM, an analog circuit architecture that implements dendritic arbors with \ac{RRAM}-based delay and weights. The delays enrich the architecture to solve temporal tasks using only feedforward connections, and are resilient to noise. DenRAM can solve the keywordspotting and hearbeat anomaly tasks with much higher accuracy, while having an equal number parameters to \ac{SRNN}.

\subsection*{Perspectives on the role of delays in temporal data processing}
Delays are at the core of the spatio-temporal processing performed by dendrites and their central importance has been demonstrated in DenRAM. We have shown that delays, and dendritic computation in general, are key to improving computational and energy efficiency of spiking neural network chips. Delays explicitly enable  the construction of temporal computational primitives and allow avoiding/reducing recurrent connections to induce temporal dynamic processing~\cite{hochreiter1998vanishing}. By providing feed-forward networks the ability to carry out spatio-temporal pattern recognition without having to use recurrent connections we obtained several benefits; Firstly, we could minimize the memory footprint and thus improve the computational efficiency; Secondly, these networks escape the vanishing and exploding gradient problems present in \acp{RNN} trained with the \ac{BPTT} algorithm. Both of these benefits are highlighted in Fig.~\ref{fig:Figure1}c, where delay-based neural networks reached superior accuracy compared to recurrent architectures on the SHD task. 
%DenRAM results are grouped together with other networks that use delays in different forms~\cite{patino_etal2023_imec_delay, hammouamri2023learning}, all fitted on the green curve. 
%The state-of-the-art \ac{SRNN} models are represented with the gray markers. 
Evidently, for the same number of parameters, delay-based models achieve higher accuracy in the \ac{SHD} task, confirming the greater computational efficiency of this type of architecture. Crucially, delay-based computation is a relatively novel concept and it is likely that the exploitation of the potential of delays has not yet been maximized. In the DenRAM architecture, similar to \cite{patino_etal2023_imec_delay}, the delay parameters are not trained and do not need to be optimized. 
Methods to adapt delays via gradient descent were recently developed~\cite{shrestha2018slayer}, but the exploration of the benefits of training delays is a recent trend \cite{sun2023adaptive, hammouamri2023learning}. 
In particular, Hammouamri et al. link the delays to temporal causal convolutions, thus making use of modern deep learning techniques to train a \ac{SNN}, and achieve the highest classification accuracy on the SHD task~\cite{hammouamri2023learning} . 

Exploiting the techniques that have been developed in deep learning to train such networks is potentially very fruitful. For example, we envision that delays can also implement temporal convolutions~\cite{oord2016wavenet}, and temporal causal Graph Neural Networks~\cite{dalgaty2023hugnet}, where past events can be grouped into a graph representation and achieve impressive results in vision tasks.

Furthermore, the hardware realization of the DenRAM architecture exhibits intriguing parallels with the concept of MLP-mixer, which has demonstrated remarkable performance in vision tasks without relying on convolution or attention mechanisms~\cite{tolstikhin2021mlpmixer}. While the MLP-mixer randomly mixes features across spatial locations and channels in a two-step process, DenRAM introduces a unique form of temporal shuffling by incorporating random delays to temporal inputs within each channel before integration.
These innovative architectural approaches, which reorganize incoming data, whether spatially (as in the case of MLP-mixer) or temporally (as in DenRAM), hold the promise of more effectively extracting complex patterns. This suggests a compelling direction for network design, potentially leading to models that are both computationally efficient and highly capable.

Grounding these novel algorithms on an energy-efficient hardware substrate is thus crucial, and hence is in the the scope of this work. Making an innovative use of the emerging memory technologies expands the functionality of the circuit as well as minimizing power consumption. 

\subsection*{Effect of heterogeneity on the performance of DenRAM}

Similar to the computational neuroscience studies on the beneficial role of heterogeneity on the performance of networks~\cite{perez_etal2021_hetero}, we have performed an ablation study on the accuracy of DenRAM on our two benchmarks, by varying the variability of the delay and weight parameters. 

\paragraph*{Heartbeat anomaly detection}

We first vary the standard deviation of the underlying normal distribution of delay RRAMs, while fixing the number of synapses per dendritic branch, and fixing the noise of the weight RRAMs to 10\%.  We find that the required mean delay, to solve the ECG task to 95\% accuracy, increases when the standard deviation of the underlying normal distribution decreases. This shows the beneficial role of heterogeneity in loosening the hardware requirements: the higher the variability of the underlying distribution, the lower the required resistance and capacitance of the delay circuit of Fig.\ref{fig:Figure2}a)~ (Supplementary Note 4).

On he other hand, noise in the programming of RRAM weights was controlled through sweeping the noise injection in terms of percent of the maximum weight in the network. The accuracy of the DenRAM was almost unaffected by noise values up to about 20\%, showing to be much more tolerant than the \ac{SRNN} architecture of iso-accuracy when no noise is injected. 

\paragraph*{Keyword spotting} 
Applying the same methodology used in the delay distribution analysis to the SHD task, we observe parallel outcomes (Supplementary Note 4). 
This consistency underscores that the characteristics we identified are not unique to a specific task but are, in fact, fundamental attributes of the DenRAM architecture. 
Furthermore, our research uncovers a significant aspect: the existence of an optimal mean delay within this framework. 

\subsection*{Delay analysis for keyword spotting}

Leveraging both spatial parameters (weights) and temporal parameters (delays), DenRAM effectively utilizes a two-dimensional approach for the separation and classification of spatio-temporal inputs. This capability is elucidated in Supplementary Note 5 and Fig. S6, where, in order to discern between input patterns within the \ac{SHD} dataset, the network strategically employs two distinct strategies. At times, it opts for ``spatial segregation'', adjusting the weighting of different channels to differentiate between patterns (e.g., when distinguishing between digits ``8'' and ``18''). 
Alternatively, it introduces ``weight dynamics'' through the delayed replications of inputs by dynamically altering the aggregated weight values of each channel over time, thereby implementing ``temporal segregation'' to distinguish between other patterns (e.g., when differentiating between digits ``8'' and ``17'').
% Melika's version
% Alternatively, it introduces ``weight dynamics'' by dynamically altering the weight values of each channel over time, thereby implementing ``temporal segregation'' to distinguish between other patterns (e.g., when differentiating between digits ``8'' and ``17'').

\subsection*{Outlook}

\paragraph*{Long time scales represented on chip} 
Using \ac{RRAM} technology, we successfully implemented delay elements that achieved periods of up to 60\,ms. This proved sufficient temporal memory for addressing the heartbeat anomaly detection task. However, it fell short in meeting the demands of the keyword spotting task, where delays of up to 500\,ms were necessary to reach a desirable accuracy. To accommodate these extended time scales using the same approach, novel technologies are required. Recent research has explored volatile resistive memory devices with tunable time scales, albeit with limitations, typically averaging in the range of 10s of ms~\cite{John_etal22, wang2019modeling,wang_etal2017_volatileRev}. 
Material engineering which results in technologies with larger time scale in their decay would prove extremely beneficial in implementing short-term dynamics and delays with minimal area on analog chips. 
On the non-volatile memory front, Ferroelectric Tunnel Junctions (FTJ)s present an opportunity for the delay element in the next generation of DenRAM, as their resistances is much higher than the RRAM devices (e.g., \ac{HRS} of FTJ can go beyond G$\Omega$ Ohms, as compared to 10s-100s of M$\Omega$ Ohms for RRAM)~\cite{fontanini_etal2022_ftj}. This is due to a different current conduction mechanism of FTJ which is based on tunneling, compared to the Ohmic conduction in RRAM. 

\paragraph*{Non-linear integration in dendrites}
Biological dendrites feature multiple characteristics and dynamical behavior that have not been accounted for in this work. For example, the non-linearity of dendritic branches \cite{acharya2022dendritic} might further enhance the computational capacity of neural networks \cite{jones2020can, chavlis2021drawing} and alone solve XOR task\cite{gidon_etal2020_dendritic}. Such investigations are the natural next step for DenRAM.

\paragraph*{On-chip learning}
The plasticity of resistive memory has not been investigated for on-chip learning in this work, despite that being a very promising aspect of such hardware substrate. Dendrites have been shown to play an important role in learning of the cortical circuits \cite{sacramento2018dendritic}, and have been previously implemented in \ac{CMOS} technology for stochastic on-chip online learning~\cite{cartiglia_rubino_etal2022_alive}. DenRAM provides an ideal architecture for implementing on-chip learning based on these concepts on RRAM based systems, which is a facet we will be exploring in future work. 

As elegantly proposed in a recent study by Boahen~\cite{boahen2022dendrocentric, boahen2023iedm}, the concept of dendrocentric computing and learning, where inputs are not only spatially weighted but also temporally considered in accordance with their arrival times, offers a compelling avenue to reduce energy consumption in the next generation of \ac{AI} hardware.
%\GI{cite also the latest work presented at IEDM from Boahen's group where they present FET devices that support the dendrocentric CD computation at the single device level: \url{https://spectrum.ieee.org/dendrites}}.
DenRAM stands as a pioneering achievement in this direction, being the first hardware implementation of dendrites that harnesses the unique properties of emerging memory technologies, particularly resistive memory devices. This marks a significant milestone in advancing dendrocentric computing and learning, setting the stage for more efficient and innovative AI hardware solutions.

\section*{Methods}

\subsection*{Design, fabrication of DenRAM circuits}
\subsubsection*{Dendritic Circuit design}
The Dendritic Circuit (Figure \ref{fig:Figure2}a) features two main sections: one devoted to generating the delay and the second to weight the output currents. The circuit takes input spikes in the form of stereotypical voltage pulses, in our design of 1.2V peak voltage and 1$\mu s$ pulse width. The pulses are applied to the IN terminal, opening the nMOS transistor. The charge on the capacitor, resting at $Vref$, is then pulled to ground during the application of the spike, causing the voltage $Vcap$ to plummet to ground. $Vref$ is set to 0.6V, a value that maximizes the dynamic range of the capacitor while reducing the read-noise of RRAM devices. As the Multiplexers' selection ($Prog$) is low, the capacitor is connected to the 1T1R Delay device, whose Bit Line is set at $Vref$, forming an $RC$ pair. During the operation of the circuit, the Delay 1T1R Word Line is open (set at 4.8V). For this reason, the capacitor's voltage $Vcap$ recharges with a time constant $\tau = RC$. A Thresholding unit detects when $Vcap$ relaxes back to the resting potential, crossing a certain threshold voltage, in our case set at 250mV. Note that the Thresholding unit is bypassed when the IN spike is applied so that only the second crossing of the threshold by the $Vcap$ potential is detected. More details on the Thresholding unit are in Supplementary Information, Fig.~2a. The output of the Thresholding unit is a spike of the same shape as the input one, just at 4.8V. This voltage pulse passes through the second Multiplexer set by the selection $Prog$ voltage, entering the second section of the Dendritic Circuit. The output spike ($V_{OUT}$), which occurs with a certain delay compared to the input spike IN, is applied to the Word Line of the Weight 1T1R. The Bit Line of the Weight RRAM is pinned at the RRAM's reading voltage (around 0.6V).
%: more details on the readout circuit in the Supplementary Information (\textcolor{red}{FM: add the readout circuit in the supplementary notes})
As a consequence, an output current $I_{OUT}$ is generated, proportional to the conductance of the Weight RRAM. This current is read out and then fed to an output LIF neuron (details in the Supplementary Information, Fig.~2b).

\subsubsection*{Fabrication/integration} 
%\textcolor{red}{@Elisa control this paragraph}
We fabricated our circuits in a 130\,nm technology, in a 200\,mm production line. The RRAM devices' stack is $TiN/Si: HfO/Ti/TiN$, formed by a $10\,$nm thick $Si:HfO$ layer sandwiched by two $4\,$nm thick $TiN$ electrodes. Notably, we selected a thicker oxide layer so that the pristine state's resistance would be maximized while the Si doping reduces the forming voltage.
%we selected a thicker oxide layer so that the Pristine State's resistance would be maximized while maintaining a low enough forming voltage. 
Each RRAM device is coupled to an access transistor, forming 1T1R structures, that are used to select RRAM devices individually during programming operations. The size of the access transistor is $650\,nm$ wide. RRAMs are built between metal layers 4 and 5, allowing to integrate them in the Back-End-of-Line, maximizing integration density. 
%Dendritic Networks featuring RRAM devices have been built in different sizes, spanning from [8 x 1] to [64 x 3] ([D, B], D being the number of dendritic circuits per branch, B being the number of branches). 
To access RRAMs in DenRAM, peripheral circuits have been designed. Each array row and column (Word Line, Bit Line, and Source Line) is interfaced by a multiplexer connecting the line either to ground or to a pad, where a programming/reading voltage could be applied. Multiplexers are operated by Shift Registers that store the addresses of the lines to be connected to the pad's programming/reading voltage. All the circuits are featured in a 200\,mm wafer and are accessed by a probe card connecting to pads of size of $[50x90]\mu m^2$ each.

\subsection*{RRAM characteristics}
Smart programming of the devices can be reached to obtain more precise conductance levels and stabilize the devices with respect to the filament relaxation resulting in a conductance shift\cite{esmanhotto_etal_2022, moro_etal2022_ISCAS}.

%\textcolor{red}{Elisa: to be written}

The resistive switching mechanisms employed in our paper's devices are based on the creation and dissolution of a conductive pathway within the device, brought about by the application of an electric field. This change in the pathway's geometry leads to distinct resistive states within the device. To execute a SET or RESET operation, we apply either a positive or negative pulse across the device, respectively. This pulse formation or disruption within the memory cell results in a decrease or increase in its resistance. When the conductive pathway is formed, the cell is in the Low Resistive State (LRS); otherwise, it is in the High Resistive State (HRS). In a SET operation, the bottom of the 1T1R structure is conventionally held at ground level, while a positive voltage is applied to the top electrode of the 1T1R. Conversely, in a RESET operation, the reverse is applied.

\subsection*{Dendritic circuit measurement setups}
%\textcolor{red}{TO BE REVIEWED FOLLOWING THE COMMENTS THAT I PUT}

The tests of the circuit involved analyzing and recording the dynamical behavior of analog CMOS circuits as well as programming and reading RRAM devices. Both phases required dedicated instrumentation, all simultaneously connected to the probe card. 
For reading the RRAM devices, Source Measure Units (SMU)s from a Keithley 4200 SCS machine were used while, for programming, a B1530A waveform generator by Keysight was used in order to send SET and RESET pulses. To maximize the stability and precision of the programming operation, SET and RESET are performed in a quasi-static manner. This means that a slow rising and falling voltage input is applied to either the Top (SET) or Bottom (RESET) electrode, while the gate is kept at a fixed value. 

To the $V_{top}(t)$, $V_{bot}(t)$ voltages, we applied a trapezoidal pulse with rising and falling times of $50\,$ns, a pulse width of $1\,$\textmu s and picked a value for $V_{gate}$ . For a SET operation, the bottom of the 1T1R structure is conventionally left at ground level, while in the RESET case the $V_{top}$ is equal to 0\,V and a positive voltage is applied to $V_{bot}$. Typical values for the SET operation are $V_{gate}$ in $[1.6-2.2]\,V$, while the $V_{top}$ peak voltage is normally at $[2-2.6]\,V$. Such values allow modulating the RRAM resistance in an interval of $[8-50]\,k\Omega$ corresponding to the \ac{LRS} of the device. For the RESET operation, the gate voltage is instead at $4.5\,V$, while the bottom electrode is reaching a peak at $[1.8-2.4]\,V$. 

The \ac{HRS} is less controllable than the \ac{LRS} due to the inherent stochasticity related to the rupture of the conductive filament, thus the HRS level is spread out in a wider $[60-1000]\,k\Omega$ interval. The reading operation is performed by limiting the $V_{top}$ voltage to $0.4\,V$, a value that avoids read disturbances, while opening the gate voltage at $4.5\,V$. 

Inputs and outputs of the dendritic circuit and the LIF neuron are analog dynamical signals. In the case of the input, we have alternated a HP 8110 pulse generator with a B1530A by Keysight combined with a B1500A Semiconductor Device Parameter Analyzer by Keysight. As a general rule, input pulses had a pulse width of $1\mu\,s$ and rise/fall time of $20\,ns$. This type of pulse is assumed as the stereotypical spiking event of a Spiking Neural Network. Concerning the outputs, a $1\,GHz$ Teledyne LeCroy oscilloscope was utilized to record the output signals of the various elements in the dendritic circuit and the LIF neuron. An Arduino Mega 2560 board for collecting statistics on read-to-read delay variability was used when testing the dendritic circuit alone, using its built-in timers for recording the delay between the input spike and the output spike, varying the timer scale in order to adapt to the delay order of magnitude and avoid overflow problems in the board registers. 
Due to the impossibility of reading resistances as high as pristine (in the order of tens to hundreds of G$\mathrm{\Omega}$), the measures of the pristine resistances are extracted from the delay measurements on 71 dies through the formula $R_i=D_i/C$.

\subsection*{\ac{RRAM}-aware noise-resilient training}
% \textcolor{red}{TO BE REVIEWED (maybe comparing it to the following part extracted from mosaic paper)}

The strategy of choice for endowing DenRAM with the ability to solve real-world tasks is hardware-aware gradient descent. The conventional backpropagation-based optimization utilized in machine learning is tailored to the hardware substrate implementing DenRAM. In particular, DenRAM makes use of RRAMs as synaptic weights: this imposes constraints on the parameters of the network, that are accounted for during the training phase. The main challenge is the stochasticity of RRAMs, resulting in overprecise weights.

We propose to address this problem by introducing the non-idealities of RRAM during the training phase. During inference, we perturb the weights of the neural network with the same variability measured on the RRAM devices. However, we apply the computed gradients to the original unperturbed weights. This methodology is similar to Quantization-Aware-Training \cite{hubara2017quantized}. Eventually, we perform a pre-training phase without RRAM's variability, so as to better initialize the hardware-aware training.

%A second method is to split the training into two phases. First, an offline part using high precision (floating point 32bits) weights. Then, a second offline part, involving an RRAM model of the resistance extracted from measurements for weights values and quantization aware techniques; that part add to the conductance measurements the zero value in order to be able to decide what are the devices to let in their pristine state when performing the online training.

%Both the phases can account for noise corresponding to the conductance shift of the RRAM due to filament relaxations \SD{(Supplementary or Fig 4?)} 

%Finally, the online training is done following a mixed-precision approach \cite{le2018mixed, demirag2021online}. This part of the training can be performed both on the hardware circuit using a computer-in-the-loop approach \textbf{[any citation here?]} and simulating it on a classic GPU. Here the weight value 0 is substituted with the LCS of the RRAM and can be reached only by weights exiting the previous phase with a non-zero value.

%To sum up, this strategy technique is able to start from a quasi hardware-unaware training to introduce smoothly RRAM non-idealities to arrive performing an online training that can be simulated thanks to its hardware friendly characteristics.

\subsection*{Heartbeat Anomaly detection task description}
% \textcolor{red}{TO BE REVIEWED}

For the Heartbeat anomaly detection task, we chose the MIT-BIH dataset \cite{moody2001}. Such a database is composed of continuous Electro-Cardio-Gram (ECG) 30-minutes recordings measured from multiple subjects. Each ECG recording is annotated by different cardiologists indicating each heartbeat as normal or abnormal, and the type of arrhythmia, when present.
The data for each subject contains the recording from two leads, but it has been demonstrated that one lead is sufficient for performing a correct classification\cite{deChazal2004}.
%since the second lead is not always the same because in the medical field the absence of standard is their standard. Such an absence of standard is reflected when a person without a medical background tries to understand the dataset: in order to make appearing complicate something which is normally easy, the labelling of the dataset is delirious (always because of the medical standard). Moreover, medical recording of data introduces noise and singularities in addition to the already present differences between patients \cite{feiteng2019}. 
%These qualities, added to the very unbalanced number of samples per class even for binary classification, make it impossible to use of all the subjects together without a heavy preprocessing: an impossible action for chips for real-time analysis of signals. 

For the \ac{ECG} task, patient 208 has been selected having the most balanced label count between normal and abnormal heartbeats of all the subjects. Labels have been grouped in normal heartbeats (labels "L", "R" and "N") and anomalies (labels "e", "j", "A", "a", "J", "S", "V", "E", "F", "/", "f" and "Q"). The 30-minute recording has been divided into segments of $180\,$time steps each around the R peak \cite{ yan2021} and divided into a train set and a test set of equal sizes. 
The input channel coming from the measurements of lead A has been converted into spikes through a sigma-delta modulator \cite{corradi2015neuromorphic} generating two different inputs for the NN: one with the so-called up spikes and the other with the down spikes.
%Finally, a complete stream of the 30-minute data input is performed, searching for the output spike positioning in time with respect to the R peak in order to check the correct real-time behavior.
This data is fed to either the DenRAM architecture or an \ac{SRNN}, and outputs are encoded as the spiking activity of the single output neuron. Large output spiking activity - above a predefined threshold - signals the detection of arrhythmia.

\subsection*{Keyword Spotting task description}
The SHD dataset~\cite{cramer_etal2020_SHD} is based on Lauscher artificial cochlea model that converts audio speech data to spike train representation with 700 channels, similar to spectrogram representation with Mel-spaced filter banks. 
It consists of 10000 recordings (8156 training, 2264 test samples) for 20 classes of spoken digits from zero to nine in both English and German languages.
We divided the original training dataset into according to a 80\%-20\% train-val split.
Each recording duration in the dataset is maximum 1.4 seconds and converted spikes are time binned into 280 5ms bins, resulting in (700 x 280) dimensionality.
We observed only 2\% of the samples have spikes after the 150th timestep, thus we truncated the input duration to 750ms. 
Only on the simulations with 256 dendritic arbors, we sampled three times along the channel dimension without overlap to obtain three augmentations with 256 channels.
This sampling improved the speed of our simulations by reducing the total number of parameters in the network and increased the both sizes of training and testing datasets.
% Loss 
The network is trained using \ac{BPTT} with the cross-entropy loss where the logits are calculated using the maximum potential over time non-linearity for each leaky-integrator output neuron.
We use a batch size of 64 for delay-based architectures and 128 for SRNN.
Each experiment reported was conducted using 3 distinct random seeds.
All experiment hyperparameters (membrane decay time constants, spike threshold, weight scaling and learning rates) are tuned separately to obtain maximum performance.

\section*{Data availability}
The MIT-BIH ECG dataset~\cite{moody2001} and the Spiking Heidelberg Datasets \cite{cramer_etal2020_SHD} dataset are publicly accessible. All other measured data are freely available upon request.

\section*{Code availability}
All software programs used in the presentation of the Article are freely available upon publication of the Article.

\section*{Acknowledgements}

We acknowledge funding support from the H2020 MeM-Scales project (871371), SNSF Starting Grant Project UNITE (TMSGI2-211461), Marie Skłodowska-Curie grant agreement No 861153, as well as European Research Council consolidator grant DIVERSE (101043854). We are grateful to Jimmy Weber for helpful discussions throughout the project.

\section*{Author contributions}

M.P. proposed the idea. M.P., F.M., E.V., and G.I. developed the dendritic circuit and network concepts. F.M. and M.P. designed and laid out the circuits for fabrication. E.V. supervised the fabrication of the circuits in hybrid RRAM-CMOS process. S.D. performed the characterizations and verification on the fabricated circuits. S.D. developed the hardware-aware training of the dendritic network for heartbeat anomaly benchmark. T.T. developed the hardware-aware training of the dendritic network for the keyword spotting task. Y.D. implemented the first delay implementation on JAX and supervised the hardware-aware simulations. All authors contributed to the writing of the manuscript. M.P. and E.V. supervised the project. 

%\section{Additional material}
%\subsection{Additional Figures}
%\begin{figure}[tb!]
%    \centering
%    \includegraphics[width=0.45\textwidth]{img/CD_separate.pdf}
%    \caption{The dendritic architecture can also be used to separate spikes by network which are not correlated but could lead to an aoutput spike or change in the membrane voltage too significant; in this example two spikes coming in with a delay of $1\,\mathrm{\mu}$s have been separated thanks to delays, without they would have led to an output spike.}
%    \label{fig:Figure5}
%\end{figure}

%\bibliographystyle{unsrtnat}
\bibliography{biblio/references,biblio/biblioncs,biblio/biblio_mosaic}  %%% Uncomment this line and comment out the ``thebibliography'' section below to use the external .bib file (using bibtex) .

\end{document}

\acrodef{AI}[AI]{Artificial Intelligence}
\acrodef{ADC}[ADC]{Analog to Digital Converter}
\acrodef{ADEXP}[AdExp-I\&F]{Adaptive-Exponential Integrate and Fire}
\acrodef{AER}[AER]{Address-Event Representation}
\acrodef{AEX}[AEX]{AER EXtension board}
\acrodef{AE}[AE]{Address-Event}
\acrodef{AFM}[AFM]{Atomic Force Microscope}
\acrodef{AGC}[AGC]{Automatic Gain Control}
\acrodef{AMDA}[AMDA]{AER Motherboard with D/A converters}
\acrodef{ANN}[ANN]{Artificial Neural Network}
\acrodef{API}[API]{Application Programming Interface}
\acrodef{ARM}[ARM]{Advanced RISC Machine}
\acrodef{ASIC}[ASIC]{Application Specific Integrated Circuit}
\acrodef{AdExp}[AdExp-IF]{Adaptive Exponential Integrate-and-Fire}
\acrodef{BCM}[BMC]{Bienenstock-Cooper-Munro}
\acrodef{BD}[BD]{Bundled Data}
\acrodef{BEOL}[BEOL]{Back-end of Line}
\acrodef{BG}[BG]{Bias Generator}
\acrodef{BMI}[BMI]{Brain-Machince Interface}
\acrodef{BPTT}[BPTT]{Backpropagation Through Time}
\acrodef{BTB}[BTB]{band-to-band tunnelling}
\acrodef{CAD}[CAD]{Computer Aided Design}
\acrodef{CAM}[CAM]{Content Addressable Memory}
\acrodef{CAVIAR}[CAVIAR]{Convolution AER Vision Architecture for Real-Time}
\acrodef{CA}[CA]{Cortical Automaton}
\acrodef{CCN}[CCN]{Cooperative and Competitive Network}
\acrodef{CDR}[CDR]{Clock-Data Recovery}
\acrodef{CFC}[CFC]{Current to Frequency Converter}
\acrodef{CHP}[CHP]{Communicating Hardware Processes}
\acrodef{CMIM}[CMIM]{Metal-insulator-metal Capacitor}
\acrodef{CML}[CML]{Current Mode Logic}
\acrodef{CMOL}[CMOL]{Hybrid CMOS nanoelectronic circuits}
\acrodef{CMOS}[CMOS]{Complementary Metal-Oxide-Semiconductor}
\acrodef{CNN}[CCN]{Convolutional Neural Network}
\acrodef{COTS}[COTS]{Commercial Off-The-Shelf}
\acrodef{CPG}[CPG]{Central Pattern Generator}
\acrodef{CPLD}[CPLD]{Complex Programmable Logic Device}
\acrodef{CPU}[CPU]{Central Processing Unit}
\acrodef{CSM}[CSM]{Cortical State Machine}
\acrodef{CSP}[CSP]{Constraint Satisfaction Problem}
\acrodef{CV}[CV]{Coefficient of Variation}
\acrodef{DAC}[DAC]{Digital to Analog Converter}
\acrodef{DAS}[DAS]{Dynamic Auditory Sensor}
\acrodef{DAVIS}[DAVIS]{Dynamic and Active Pixel Vision Sensor}
\acrodef{DBN}[DBN]{Deep Belief Network}
\acrodef{DFA}[DFA]{Deterministic Finite Automaton}
\acrodef{DIBL}[DIBL]{drain-induced-barrier-lowering}
\acrodef{DI}[DI]{delay insensitive}
\acrodef{DMA}[DMA]{Direct Memory Access}
\acrodef{DNF}[DNF]{Dynamic Neural Field}
\acrodef{DNN}[DNN]{Deep Neural Network}
\acrodef{DOF}[DOF]{Degrees of Freedom}
\acrodef{DPE}[DPE]{Dynamic Parameter Estimation}
\acrodef{DPI}[DPI]{Differential Pair Integrator}
\acrodef{DRRZ}[DR-RZ]{Dual-Rail Return-to-Zero}
\acrodef{DRAM}[DRAM]{Dynamic Random Access Memory}
\acrodef{DR}[DR]{Dual Rail}
\acrodef{DSP}[DSP]{Digital Signal Processor}
\acrodef{DVS}[DVS]{Dynamic Vision Sensor}
\acrodef{DYNAP}[DYNAP]{Dynamic Neuromorphic Asynchronous Processor}
\acrodef{EBL}[EBL]{Electron Beam Lithography}
\acrodef{EDVAC}[EDVAC]{Electronic Discrete Variable Automatic Computer}
\acrodef{EEG}[EEG]{Electroencephalography}
\acrodef{ECG}[ECG]{Electrocardiography}
\acrodef{EMG}[EMG]{Electromyography}
\acrodef{EIN}[EIN]{Excitatory-Inhibitory Network}
\acrodef{EM}[EM]{Expectation Maximization}
\acrodef{EPSC}[EPSC]{Excitatory Post-Synaptic Current}
\acrodef{EPSP}[EPSP]{Excitatory Post-Synaptic Potential}
\acrodef{ESN}[ESN]{Echo state Network }
\acrodef{EZ}[EZ]{Epileptogenic Zone}
\acrodef{FDSOI}[FDSOI]{Fully-Depleted Silicon on Insulator}
\acrodef{FET}[FET]{Field-Effect Transistor}
\acrodef{FFT}[FFT]{Fast Fourier Transform}
\acrodef{FI}[F-I]{Frequency-Current}
\acrodef{FPGA}[FPGA]{Field Programmable Gate Array}
\acrodef{FR}[FR]{Fast Ripple}
\acrodef{FSA}[FSA]{Finite State Automaton}
\acrodef{FSM}[FSM]{Finite State Machine}
\acrodef{GIDL}[GIDL]{gate-induced-drain-leakage}
\acrodef{GOPS}[GOPS]{Giga-Operations per Second}
\acrodef{GPU}[GPU]{Graphical Processing Unit}
\acrodef{GUI}[GUI]{Graphical User Interface}
\acrodef{HAL}[HAL]{Hardware Abstraction Layer}
\acrodef{HFO}[HFO]{High Frequency Oscillation}
\acrodef{HH}[H\&H]{Hodgkin \& Huxley}
\acrodef{HMM}[HMM]{Hidden Markov Model}
\acrodef{HCS}[HCS]{High-Conductive State}
\acrodef{HRS}[HRS]{High-Resistive State}
\acrodef{HR}[HR]{Human Readable}
\acrodef{HSE}[HSE]{Handshaking Expansion}
\acrodef{HW}[HW]{Hardware}
\acrodef{ICT}[ICT]{Information and Communication Technology}
\acrodef{IC}[IC]{Integrated Circuit}
\acrodef{IEEG}[iEEG]{intracranial electroencephalography}
\acrodef{IF2DWTA}[IF2DWTA]{Integrate \& Fire 2--Dimensional WTA}
\acrodef{IFSLWTA}[IFSLWTA]{Integrate \& Fire Stop Learning WTA}
\acrodef{IF}[I\&F]{Integrate-and-Fire}
\acrodef{IMU}[IMU]{Inertial Measurement Unit}
\acrodef{INCF}[INCF]{International Neuroinformatics Coordinating Facility}
\acrodef{INI}[INI]{Institute of Neuroinformatics}
\acrodef{IO}[I/O]{Input/Output}
\acrodef{IPSC}[IPSC]{Inhibitory Post-Synaptic Current}
\acrodef{IPSP}[IPSP]{Inhibitory Post-Synaptic Potential}
\acrodef{IP}[IP]{Intellectual Property}
\acrodef{ISI}[ISI]{Inter-Spike Interval}
\acrodef{IoT}[IoT]{Internet of Things}
\acrodef{JFLAP}[JFLAP]{Java - Formal Languages and Automata Package}
\acrodef{LEDR}[LEDR]{Level-Encoded Dual-Rail}
\acrodef{LFP}[LFP]{Local Field Potential}
\acrodef{LIF}[LIF]{Leaky Integrate and Fire}
\acrodef{LLC}[LLC]{Low Leakage Cell}
\acrodef{LNA}[LNA]{Low-Noise Amplifier}
\acrodef{LPF}[LPF]{Low Pass Filter}
\acrodef{LCS}[LCS]{Low-Conductive State}
\acrodef{LRS}[LRS]{Low-Resistive State}
\acrodef{LSM}[LSM]{Liquid State Machine}
\acrodef{LTD}[LTD]{Long Term Depression}
\acrodef{LTI}[LTI]{Linear Time-Invariant}
\acrodef{LTP}[LTP]{Long Term Potentiation}
\acrodef{LTU}[LTU]{Linear Threshold Unit}
\acrodef{LUT}[LUT]{Look-Up Table}
\acrodef{LVDS}[LVDS]{Low Voltage Differential Signaling}
\acrodef{MCMC}[MCMC]{Markov-Chain Monte Carlo}
\acrodef{MEMS}[MEMS]{Micro Electro Mechanical System}
\acrodef{MFR}[MFR]{Mean Firing Rate}
\acrodef{MIM}[MIM]{Metal Insulator Metal}
\acrodef{MLP}[MLP]{Multilayer Perceptron}
\acrodef{MOSCAP}[MOSCAP]{Metal Oxide Semiconductor Capacitor}
\acrodef{MOSFET}[MOSFET]{Metal Oxide Semiconductor Field-Effect Transistor}
\acrodef{MOS}[MOS]{Metal Oxide Semiconductor}
\acrodef{MRI}[MRI]{Magnetic Resonance Imaging}
\acrodef{NDFSM}[NDFSM]{Non-deterministic Finite State Machine} 
\acrodef{ND}[ND]{Noise-Driven}
\acrodef{NEF}[NEF]{Neural Engineering Framework}
\acrodef{NHML}[NHML]{Neuromorphic Hardware Mark-up Language}
\acrodef{NIL}[NIL]{Nano-Imprint Lithography}
\acrodef{NMDA}[NMDA]{N-Methyl-D-Aspartate}
\acrodef{NME}[NE]{Neuromorphic Engineering}
\acrodef{NN}[NN]{Neural Network}
\acrodef{NRZ}[NRZ]{Non-Return-to-Zero}
\acrodef{NSM}[NSM]{Neural State Machine}
\acrodef{OR}[OR]{Operating Room}
\acrodef{OTA}[OTA]{Operational Transconductance Amplifier}
\acrodef{PCB}[PCB]{Printed Circuit Board}
\acrodef{PCHB}[PCHB]{Pre-Charge Half-Buffer}
\acrodef{PCM}[PCM]{Phase Change Memory}
\acrodef{PE}[PE]{Processing Element}
\acrodef{PFA}[PFA]{Probabilistic Finite Automaton}
\acrodef{PFC}[PFC]{prefrontal cortex}
\acrodef{PFM}[PFM]{Pulse Frequency Modulation}
\acrodef{PR}[PR]{Production Rule}
\acrodef{PSC}[PSC]{Post-Synaptic Current}
\acrodef{PSP}[PSP]{Post-Synaptic Potential}
\acrodef{PSTH}[PSTH]{Peri-Stimulus Time Histogram}
\acrodef{QDI}[QDI]{Quasi Delay Insensitive}
\acrodef{RAM}[RAM]{Random Access Memory}
\acrodef{RDF}[RDF]{random dopant fluctuation}
\acrodef{RELU}[ReLu]{Rectified Linear Unit}
\acrodef{RLS}[RLS]{Recursive Least-Squares}
\acrodef{RMSE}[RMSE]{Root Mean Squared-Error}
\acrodef{RMS}[RMS]{Root Mean Squared}
\acrodef{RNN}[RNN]{Recurrent Neural Networks}
\acrodef{RRAM}[RRAM]{Resistive Random Access Memory}
\acrodef{RSNN}[RSNN]{Recurrent Spiking Neural Network}
\acrodef{ROLLS}[ROLLS]{Reconfigurable On-Line Learning Spiking}
\acrodef{RRAM}[RRAM]{Resistive Random Access Memory}
\acrodef{R}[R]{Ripples}
\acrodef{SAC}[SAC]{Selective Attention Chip}
\acrodef{SAT}[SAT]{Boolean Satisfiability Problem}
\acrodef{SCX}[SCX]{Silicon CorteX}
\acrodef{SD}[SD]{Signal-Driven}
\acrodef{SEM}[SEM]{Spike-based Expectation Maximization}
\acrodef{SLAM}[SLAM]{Simultaneous Localization and Mapping}
\acrodef{SNN}[SNN]{Spiking Neural Network}
\acrodef{SNR}[SNR]{Signal to Noise Ratio}
\acrodef{SOC}[SOC]{System-On-Chip}
\acrodef{SOI}[SOI]{Silicon on Insulator}
\acrodef{SP}[SP]{Separation Property}
\acrodef{SHD}[SHD]{Spiking Heidelberg Digit}
\acrodef{SRAM}[SRAM]{Static Random Access Memory}
\acrodef{SRNN}[SRNN]{Spiking Recurrent Neural Network}
\acrodef{STDP}[STDP]{Spike-Timing Dependent Plasticity}
\acrodef{STD}[STD]{Short-Term Depression}
\acrodef{STP}[STP]{Short-Term Plasticity}
\acrodef{STT-MRAM}[STT-MRAM]{Spin-Transfer Torque Magnetic Random Access Memory}
\acrodef{STT}[STT]{Spin-Transfer Torque}
\acrodef{SW}[SW]{Software}
\acrodef{TCAM}[TCAM]{Ternary Content-Addressable Memory}
\acrodef{TFT}[TFT]{Thin Film Transistor}
\acrodef{TPU}[TPU]{Tensor Processing Unit}
\acrodef{USB}[USB]{Universal Serial Bus}
\acrodef{VHDL}[VHDL]{VHSIC Hardware Description Language}
\acrodef{VLSI}[VLSI]{Very Large Scale Integration}
\acrodef{VOR}[VOR]{Vestibulo-Ocular Reflex}
\acrodef{WCST}[WCST]{Wisconsin Card Sorting Test}
\acrodef{WTA}[WTA]{Winner-Take-All}
\acrodef{XML}[XML]{eXtensible Mark-up Language}
\acrodef{CTXCTL}[CTXCTL]{CortexControl}
\acrodef{divmod3}[DIVMOD3]{divisibility of a number by three}
\acrodef{hWTA}[hWTA]{hard Winner-Take-All}
\acrodef{sWTA}[sWTA]{soft Winner-Take-All}
\acrodef{APMOM}[APMOM]{Alternate Polarity Metal On Metal}
\acrodef{SRNN}[SRNN]{Spiking Recurrent Neural Networks}
\acrodef{fMRI}[fMRI]{functional Magnetic Resonance Imaging}
\acrodef{RL}[RL]{Reinforcement Learning}
\acrodef{ES}[ES]{Evolutionary Strategies}
\acrodef{SL}[SL]{Source Line}
\acrodef{BL}[BL]{Bit Line}
\acrodef{WL}[WL]{Word Line}
\acrodef{CD}[CD]{Coincidence Detection}
\acrodef{NMDA}[NMDA]{N-methyl-D-aspartate}

%\maketitle

\begin{center}
\textbf{\large Supplemental Materials:\\ DenRAM: Neuromorphic Dendritic Architecture with RRAM for Efficient Temporal Processing with Delays}
\end{center}

\vspace{1cm}

\subsection*{Supplementary Note 1}
\begin{figure}[ht!]
    \centering
    \includegraphics[width=0.6\textwidth]{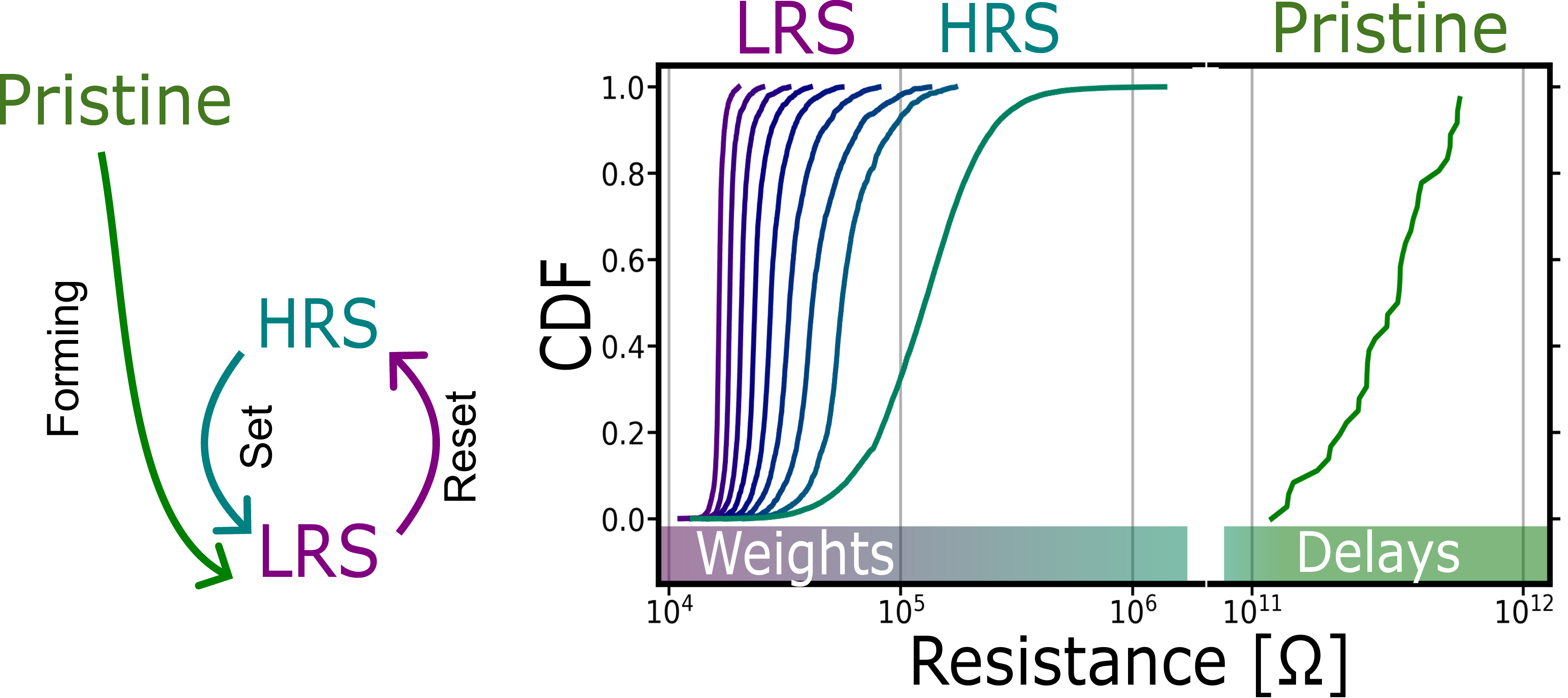}
    \caption{Characterization of the RRAM device and link to their role in DenRAM. Initially, RRAM devices are in the Pristine State. With a Forming operation, they are programmed in the \ac{LRS}. With a Reset operation, they are programmed to the \ac{HRS}. A Set operation programs the device back to the LRS. Delay RRAMs are left in the Pristine state, to maximize delays. Weight RRAMs are programmed in LRS or HRS.}
    \label{fig:sup_rram}
\end{figure}

We performed a thorough electrical characterization of a 16kbit RRAM array, with devices identical to the ones found in DenRAM. Devices are initially in the Pristine State, exhibiting large resistance, on the order of $G\Omega$. With a Forming operation - application of a positive voltage (>3V) to the Bit Line while the Source Line at ground - a conductive filament is formed in the device, programming the RRAM in the \ac{LRS}. We performed an adaptive Forming operation, meaning that we applied sequentially higher voltage pulses of the duration of 1$\mu s$ until all the devices were formed. With a Reset operation - Bit Line at ground, Source Line with a positive voltage (2.3V) the conductive filament is broken, programming the RRAMs in the \ac{HRS}. With a Set operation - Bit Line at a positive voltage [1.8-2.3]V, Source Line at ground - the devices can be programmed at the LRS. Varying the voltage applied on the Word Line [1.6-2.2]V during programming, different resistivity can be achieved, as shown in Figure~\ref{fig:sup_rram}.

\subsection*{Supplementary Note 2}
\begin{figure}[ht!]
    \centering
    \includegraphics[width=0.8\textwidth]{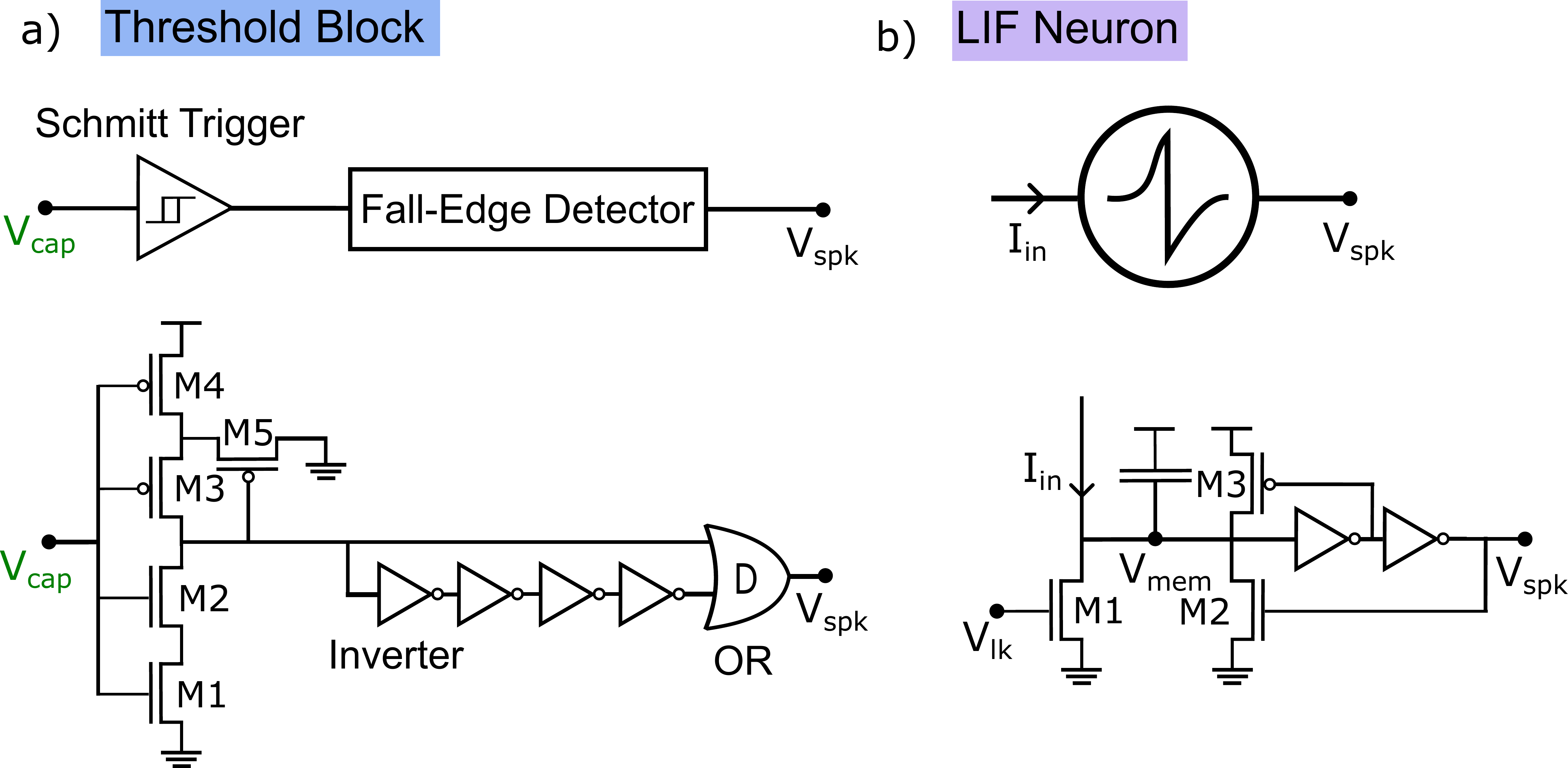}
    \caption{ Circuits blocks of the DenRAM architecture. a) The Threshold Blok in the Dendritic Circuit features two sub-blocks: a Schmitt Trigger and a Fall-Edge-Detector. The schematics of the two sub-blocks are shown below. b) Leaky-Integrate-Fire neuron circuit. The circuit features a Capacitor, a biased transistor M1 controlling the leakage rate with the $V_{lk}$ bias voltage, and an output section to produce the output spikes ($V_{spk}$). Transistor M2 resets the membrane voltage Vmem after a spike, and M3 provides positive feedback to avoid meta-stability when producing an output spike.}
    \label{fig:sup_thr}
\end{figure}

We introduced the dendritic circuit, the building block of the DenRAM architecture. An important component in this circuit is the Thresholding block, highlighted in Figure \ref{fig:sup_thr}a, featuring a Schmitt Trigger and a Fall-Edge-Detector. The Schmitt Trigger circuit sets a threshold to the voltage at the Capacitor in the dendritic circuit, which relaxes from ground back to $V_{ref}$. The double threshold of the Schmitt Trigger allows for minimizing static power consumption. At the detection of the threshold, corresponding to the delay time produced by the Delay RRAM, the output of the Schmitt Trigger flips from high (1.2V) to low (0V). This triggers the Fall-Edge-Detector, built by a chain of Inverters and an OR circuit. The Fall-Edge-Detector produces the spike that is then fed to the output section of the dendritic circuit.\\
Another important element of DenRAM is the Leaky-Integrate-and-Fire neuron circuit (Fig.~\ref{fig:sup_thr}b). We design a simple LIF neuron taking an input current, and accumulating it on a membrane capacitor, thus increasing the membrane voltage $V_{mem}$. Transistor M1 is used to leak some of the current accumulated on the capacitor, and its bias $Vlk$ can be tuned to control the leakage rate. An output section features a chain of inverters that detect the overcoming of the threshold by the membrane voltage $V_{mem}$. Such threshold crossing is favored by the positive feedback circuit involving transistor M3, which avoids the meta-stability of the membrane voltage. Transistor M2 is activated when the threshold voltage of the 2 inverters is crossed, grounding the voltage at the capacitor, and resetting the LIF neuron. Such a circuit is also assumed as the neuron model for the SRNNs system-level simulations.

\subsection*{Supplementary Note 3}
DenRAM architecture leverages spatio-temporal features of the inputs to perform computation, implementing coincidence detection (CD) as illustrated in Fig.~3c. To perform CD, temporally coincident spikes are assigned high weight, by setting the Weight RRAM to the \ac{LRS}. However, it is also able to do the opposite: to set the weight of the dendritic circuit in HRS, so that coincidence is not detected. 

\begin{figure}[hb!]
    \centering
    \includegraphics[width=0.50\textwidth]{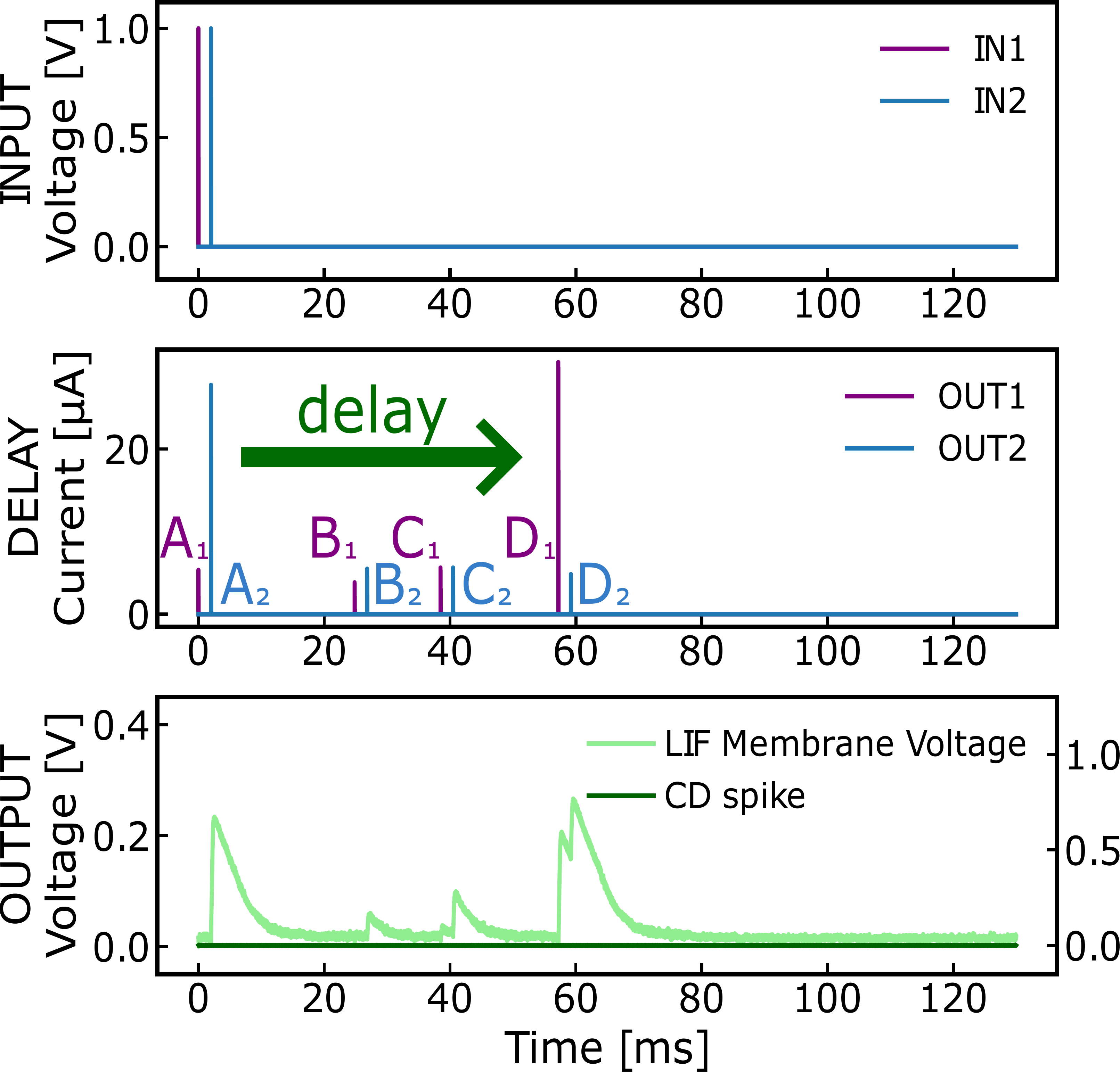}
    \caption{The dendritic architecture can also be used to separate spikes by the network that are not correlated but could lead to an output spike or change in the membrane voltage too significant; in this example, two spikes coming in with a delay of $1\,\mathrm{\mu}$s have been separated thanks to delays, without they would have led to an output spike.}
    \label{fig:sup1}
\end{figure}

Figure~\ref{fig:sup1} shows how two spikes arriving very close in time can be separated by modulating the weight of the delays associated with the input of the two channels. Differently that in the experiment shown in Figure~3c, the weight associated with spike $D_1$ is set to \ac{HRS}. In this way, the effect of the pair of coincident spikes $D_1, D_2$ on the output membrane voltage is lower and it prevents the neuron from spiking and thus detecting coincidence.

This mechanism can be critical if a group of spikes has to be correlated with another one without considering the influence of some previous spikes: in this case, CD and spike separation will be performed together by the network shifting one channel in time.

\subsection*{Supplementary Note 4}
We performed a thorough analysis of the ECG task, varying different parameters regarding the DenRAM architecture. This study, for such a simple architecture, involves the variation of two parameters: the standard deviation of the distribution used for sampling the delays and the number of delays per dendrites (i.e. the number of synapses).

\begin{figure}[ht!]
    \centering
    \includegraphics[width=0.85\textwidth]{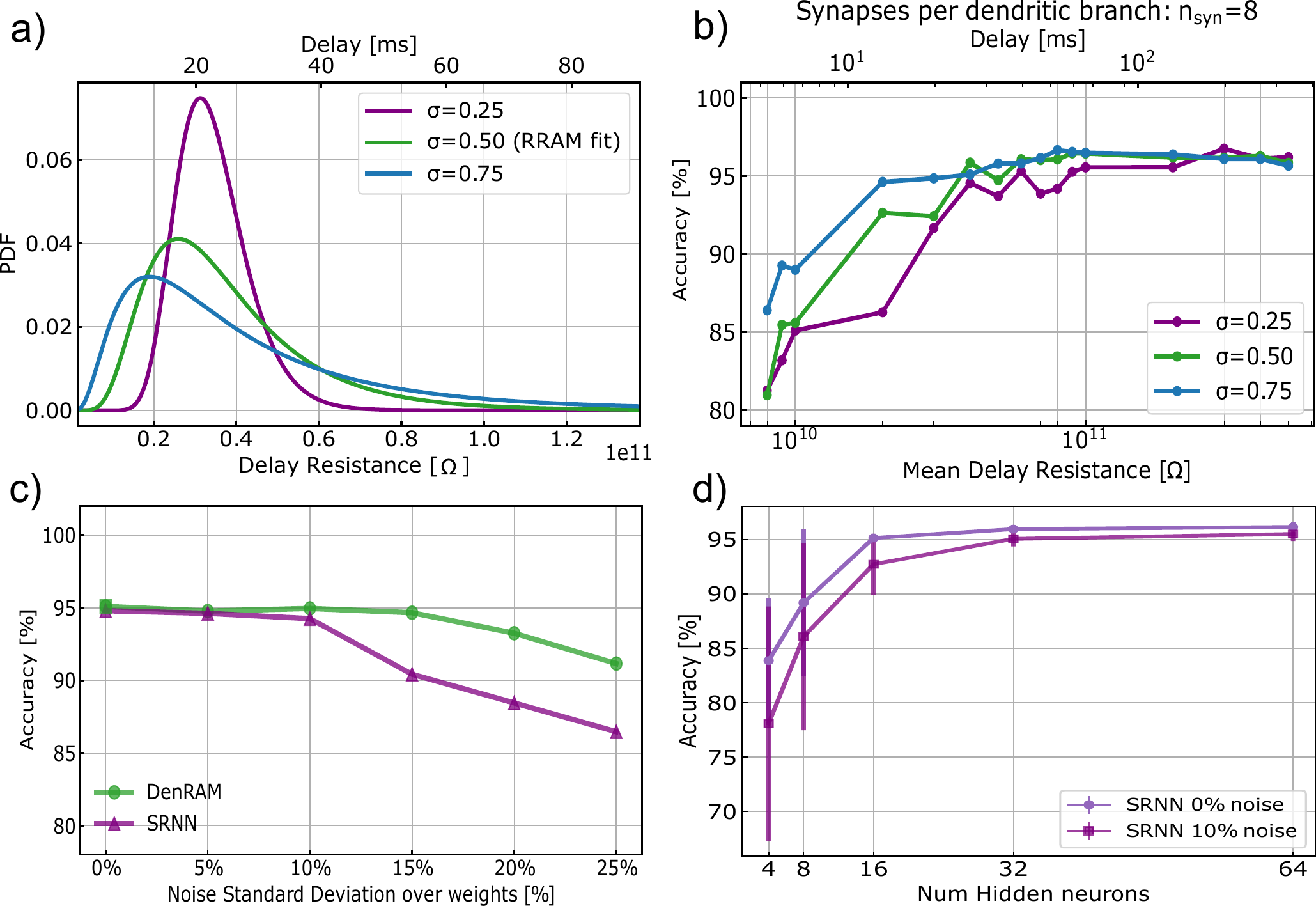}
    \caption{ Ablation study on DenRAM applied on the ECG task. a) Lognormal distribution of the Delay RRAM ($R_d$). We denote $\sigma$ as the standard deviation of the underlying normal distribution of RRAMs. The RRAM data are fit with $\sigma=0.5$. b) Accuracy as a function of the mean of the Delay RRAM's distribution, for different distributions. Weights are subject to 10\% noise. c) Accuracy of DenRAM and the SRNN as a function of the noise added on the weights, representing the variability of the Weight RRAMs. d) impact of the number of hidden neurons on the performance of the SRNN, with and without RRAM-calibrated noise applied to the weights.}
    \label{fig:sup4}
\end{figure}

It is interesting to analyze the effect of varying the distribution of the Delay RRAM ($R_d$) on the performance of DenRAM. Figure \ref{fig:sup4}a shows the fit with a lognormal distribution of the Delay RRAM data ($\sigma=0.5$), as well as how such distribution would look if the standard deviation would change.
Based on these distributions, we performed system-level simulation on the ECG task fixing the number of synapses per dendritic branch in DenRAM. In Figure.~\ref{fig:sup4}b, we sweep the mean of the delay distribution with three standard deviation values, analyzing performance. Again, the green curve ($\sigma=0.5$) is calibrated on hardware data, but the plot shows that a standard deviation of 0.75 is more beneficial when the mean of the delays is low. It is probably because of the long tail of the lognormal distribution with $\sigma=0.75$, guaranteeing long delays despite a low average. Nonetheless, RRAM-calibrated DenRAM shows high accuracy (>95\%) when the mean delay is higher than 20~ms.\\
Variability in the Weight RRAM \cite{esmanhotto_etal_2022} is also impacting the performance of DenRAM. As explained in the Method section, we apply noise-resilient training by applying a Gaussian noise during learning \cite{moro_medium_2023}. We investigate the performance of DenRAM and an SRNN performing at the same level (95\%) with no variability in the weights. The Standard Deviation of the noise added to the weight is normalized by the largest weight in the layer where Gaussian noise is applied. Increasing the noise level in the DenRAM results in a modest decrease in performance, that only noticeable when the noise standard deviation is larger than 15\%. In the SRNN instead, noise begins to decrease performance when the standard deviation is at 10\%, meaning that DenRAM is more noise-resilient to the variability of weights. 
We also investigate the impact of the number of hidden neurons on the performance of the SRNN. Increasing the number of hidden neurons increases the memory footprint, but it also improves classification accuracy and resilience to synaptic variability. In Figure~\ref{fig:sup4}d, we show the classification accuracy of the SRNN with either a 0\% noise (ideal weights) or 10\% noise applied to the weights. An SRNN with 32 hidden neurons is required to solve the task and match the performance of the DenRAM architecture. This network size is thus assumed for the Memory Footprint and Power Consumption comparisons in Figure~4b,c.

\begin{figure}[ht!]
    \centering
    \includegraphics[width=.85\textwidth]{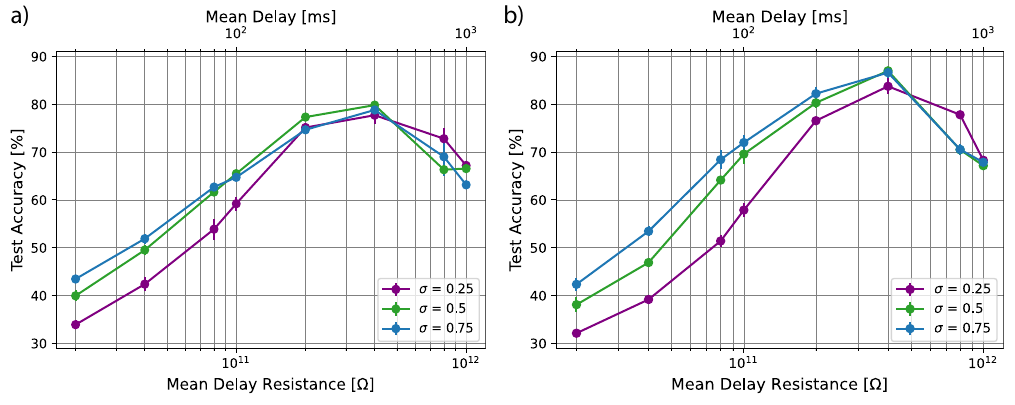}
    \caption{Ablation study on two DenRAM architectures applied on the SHD task. a) Accuracy as a function of the mean of the Delay RRAM's distribution, for different distributions, using 256 inputs and 8 delays. b) Accuracy as a function of the mean of the Delay RRAM's distribution, for different distributions, using 700 inputs and 16 delays. We denote $\sigma$ as the standard deviation of the underlying normal distribution of RRAMs. Weights are subject to 10\% noise.}
    \label{fig:supp_shd_as}
\end{figure}

We carry out a similar analysis of the SHD task using two networks with different number of inputs and different number of delays. (Fig. \ref{fig:supp_shd_as}.).
We first see that for low mean resistance, higher standard deviation helps the network to reach higher accuracy. 
For high mean resistance, the standard deviation importance is less clear. 
This behavior can be explained by the sampling of the log-normal, which is generated by sampling a normal distribution and then taking the exponential of the sampled value. 
Thus increasing the mean of the log-normal distribution while keeping the standard deviation of the underlying normal distribution will actually yield a distribution with increased standard deviation. 
Finally we see that the accuracy increases with the mean delay until 400 ms, past this value, the accuracy decreases, implying that there's an optimal mean delay in the range of 200 ms to 700ms.

%\subsection*{Supplementary Note 5}
%When dealing with RRAM technology for hardware-implemented NNs, it is important to consider the variability in the resistance, but not only. Indeed, the other phenomenon taking place after the device is programmed is filament relaxation.

%Even if intricated programming solutions have been proposed \cite{esmanhotto_etal_2022}, they can suffer from problems due to limited OxRAM endurance and trade-off between programming time and resistance precision. This motivates us to generate a variability-resilient algorithm accounting for this kind of noise \cite{moro_medium_2023}.

%We carry out an analysis of the effect of noise on the weight of DenRAM and the SRNN.
%The noise is injected during the training through two techniques: considering half of the quantity used during the inference phase, or just not considered during the training. The quantity of noise injected is always referred to as the maximum conductance obtainable with the devices.
%Fig \ref{fig:sup3} illustrates the results of the analysis: DenRAM is not only better than the SRNN but also more resistant to noise showing a loss in accuracy negligible up to a noise injected corresponding to 10\% of the maximum conductance.

%\begin{figure}[ht!]
%    \centering
%    \includegraphics[width=0.45\textwidth]{img/sup3.pdf}
%    \caption{Noise injection in DenRAM and SRNN with 100 neurons, both considering the case in which the noise is considered during training and the case where noise is injected only at inference.}
%    \label{fig:sup3}
%\end{figure}

\subsection*{Supplementary Note 5}
\begin{figure}[ht!]
    \centering
    \includegraphics[width=1\textwidth]{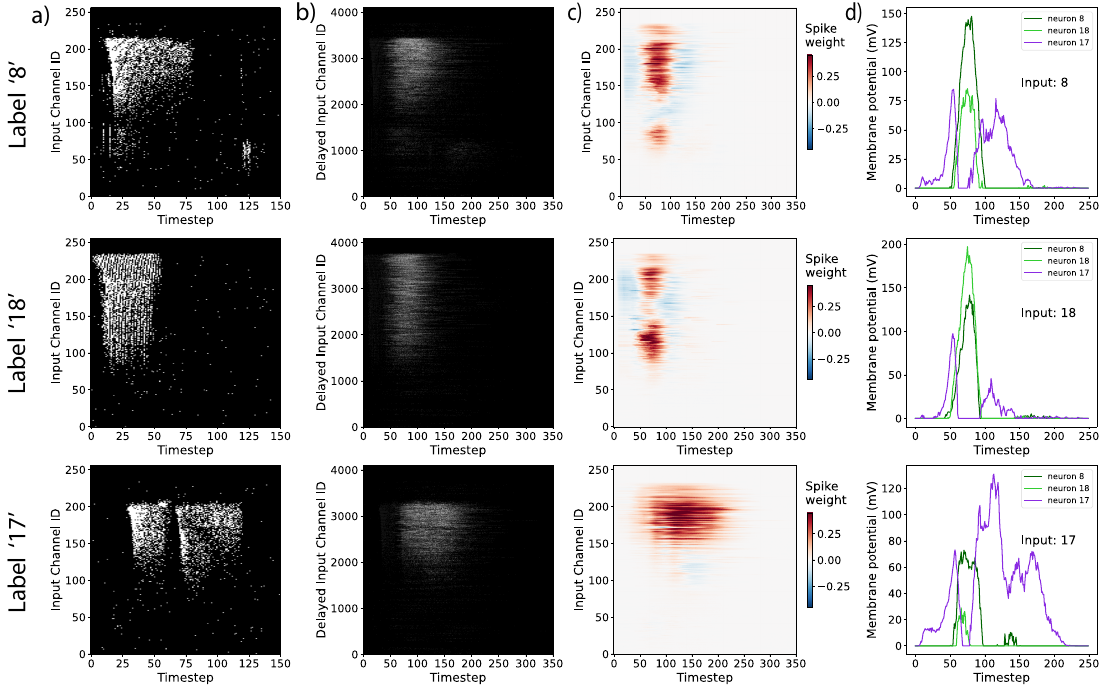}
    \caption{Analysis of the weights and delays for the D2 network (DenRAM with 16 delays per channel, using 256 input channels) . We focus on three classes from the \ac{SHD} dataset: spoken digit ``8'', ``18'' and ``17''. a) Raster plots representing individual training samples for three classes. These original samples comprise 256 channels sampled from the 700 input channels from the original dataset, and span 150 timesteps. b) Raster plots illustrating time-delayed versions of the same samples after feeding into DenRAM. The DenRAM architecture generates time-shifted replicas of the original inputs, expanding them to 4096 channels (256 channels $\times$ 16 delays per channel) and 350 timesteps. c) After learning, for each channel, the 16 delayed version are weighted and summed, giving rise to an aggregated representation of weighted-delayed input for each channel, condensing back the 4096 channels to 256. Red-shifted color represent  positive weights, and blue-shifted colors represent negative ones. The network learns to weight each aggregate-channel differently at different time steps. These plots present the weighting of the average spike train across all samples from the respective classes. d) Evolution of the output membrane potentials of the three classes when exposed to the respective inputs.}
    \label{fig:supp_shd}
\end{figure}

%\TT{TO CONTINUE}
%The DenRAM  architecture introduced in this paper replicates the input spike trains (see Fig. \ref{fig:supp_shd}a.) $n_{delays}$ times, randomly shifting these replicas in time (refer to Fig. \ref{fig:supp_shd}.). 
As explained in the main text, the DenRAM architecture delays each input channel with a population of $n_{delays}$, randomly sampled from a fixed log-normal distribution which models the measurements shown in Fig. 2d. 
Here, we are applying the inputs from the \ac{SHD} dataset to the DenRAM architecture. Fig.~\ref{fig:supp_shd}a shows  sub-sampled version of the  digits ``8'', ``18'', and ``17'' of the \ac{SHD} dataset, from 700 input channels to 256 chanels. Each of the channels in these examples are delayed $n_{delays}=16$ times in the case shown in the figure. 
Consequently, the output neurons receive $n_{in} \times n_{delays}$ spike trains (in this case $256 \times 16=4096$ delayed inputs, shown in Fig.~\ref{fig:supp_shd}b. 
After we train the network, it learns to weigh each of these 4096 channels in order to classify the presented input. To analyze the weight-delay pair, we do a weighted average of the $n_{delays}$ delayed channels for each input channel, giving rise to an aggregate representation in Fig.~\ref{fig:supp_shd}c. 
%This process becomes clearer when we aggregate these replicated spike trains, corresponding to the same input, into a single train. By applying the corresponding weights to each copy and then summing them, 
We observe that each input channel acquires a time-varying weight (illustrated in Fig. \ref{fig:supp_shd}c) to solve the task. 
For instance in digit ``8'', channels 150 to 200 exhibit a dynamic pattern, initially showing inhibition, then strong excitation, and finally returning to inhibition. 
Although digits ``8'' and ``18'' display similar patterns when delayed (see Fig. \ref{fig:supp_shd}c), the network distinguishes them by utilizing different input channels, a phenomenon we term 'spatial segregation'. 
In the case of digit ``17'', the selected input channels resemble those for digit ``8'', yet the delays cover a more extended time range, indicating 'temporal segregation'. 
This suggests that while certain input channels are shared across different digits, the network differentiates them through the temporal distribution of these channels' activations. 

This dual strategy of spatial and temporal segregation enables the network to effectively distinguish between various inputs, even when they exhibit similar delayed signatures. 
In the illustration provided in Fig. \ref{fig:supp_shd}d, the dynamic behavior of the output membrane voltage for the selected digits substantiates our observations. 
In the case of spatial segregation, it is evident that while the membrane voltage peaks for digits ``8'' and ``18'' occur simultaneously, their magnitudes are influenced by the specific inputs, which helps with their segregation. 
For example, in the case of digit ``18'' as the input, activation of channels 75 to 125 plays a key role to reduce the potentiation of neuron ``8'', whose corresponding weights are negative, while increasing the potentiation of neuron ``18'', whose correspoinding weights are positive. Therefore, the network correctly gives the maximum membrane potential to output neuron ``18'', classifying digit `18'. 

In the case of temporal segregation, when digit ``17'' is fed as the input, the variation in weight dynamics becomes a critical distinguishing factor. In this scenario, the response pattern for neuron ``8'' transitions from inhibition to excitation and back to inhibition, whereas for neuron ``17'', the pattern is consistently excitatory, leading to a higher peak in its membrane potential, correctly classifying the input.
%In Fig. \ref{fig:supp_shd}d), the output membrane voltage dynamics of the three focused digits confirm this claim. We can see that the pics of voltages of digits 8 and 18 coincide in time but their amplitude depend on the input. More precisely when feeding a 18, which excites channels 75 to 125, the neuron 8 gets less potentiated as its mapping give less importance to this channel. When feeding a 17, the weight dynamics is the main determinant, as the mapping of 8 is inhibiting then exciting then inhibiting again, while the mapping of 17 is only exciting which makes it reach a higher pic potential.

\bibliographystyle{unsrt}
\bibliography{biblio/references,biblio/biblioncs,biblio/biblio_mosaic}